\documentclass[onecolumn,amsmath,amssymb,nofootinbib,12pt]{article}
\pdfoutput=1
\usepackage{jheppub}

\usepackage[font={small}]{caption}
\usepackage{tikz}
\usepackage{subcaption}
\usepackage{graphicx}
\usepackage{dcolumn}
\usepackage{float}
\usetikzlibrary{fadings}
\usetikzlibrary{decorations.pathmorphing}
\usetikzlibrary{decorations.pathreplacing,decorations.markings}
\usetikzlibrary{intersections}
\tikzset{snake it/.style={decorate, decoration=snake}}
\usepackage{mathrsfs} 
\usepackage{fancyhdr}
\usetikzlibrary{shapes,arrows,chains}
\usetikzlibrary[calc]
\tikzfading[name=fade out, inner color=transparent!0,
  outer color=transparent!100]
\usetikzlibrary{decorations.pathmorphing}
\tikzset{snake it/.style={decorate, decoration=snake}}
\usepackage{multicol}
\usepackage{cancel}
\usepackage{mathtools}
\usepackage{hyperref}
\def\({\left(} \def\){\right)}
\def\[{\left[} \def\]{\right]}

\newcommand{\ie}{{\it i.e.,}\ }

\makeatletter
\def\@fpheader{\relax}
\makeatother

\newcommand{\be}{\begin{equation}}
\newcommand{\ee}{\end{equation}}
\newcommand{\bea}{\begin{eqnarray}}
\newcommand{\eea}{\end{eqnarray}}

\renewcommand{\eqref}[1]{(\ref{#1})}

\numberwithin{equation}{section}

\begin{document}


%
\title{Traversable wormholes in AdS and bounds on information transfer}
\author{Ben Freivogel$^{1,2}$, Dami\'an A. Galante$^1$, Dora Nikolakopoulou$^1$ and Antonio Rotundo$^1$}

\affiliation{$^1$ Institute for Theoretical Physics Amsterdam and Delta Institute for Theoretical Physics,}
\affiliation{$^2$ GRAPPA, \\
University of Amsterdam, Science Park 904, 1090 GL Amsterdam, the Netherlands.}

\emailAdd{benfreivogel@gmail.com}
\emailAdd{d.a.galante@uva.nl}
\emailAdd{t.nikolakopoulou@uva.nl}
\emailAdd{af.rotundo@gmail.com}

%
%
%
%
%
%
%
%
%
\abstract{We analyze the amount of information that can be sent through the traversable wormholes of Gao, Jafferis, and Wall. Although we find that the wormhole is open for a proper time shorter than the Planck time,  the transmission of a signal through the wormhole can  sometimes remain within the semiclassical regime. For black holes with horizons of order the AdS radius, information cannot be reliably sent through the wormhole. However, black holes with horizon radius much larger than the AdS radius do allow for the transmission of a number of quanta of order the horizon area in AdS units. 
More information can be sent through the wormhole by increasing the number of light fields contributing to the negative energy. 
Our bulk computations agree with a boundary analysis based on quantum teleportation.}

%
%
\maketitle


\section{Introduction}

\

Physicists and non-physicists  have speculated about the possibility of connecting distant pieces of spacetime by creating a ``shortcut'' joining them \cite{Morris:1988cz}.  A connection observers could travel through, called a traversable wormhole, remained in the realm of science fiction until a few years ago, when  Gao, Jafferis and Wall (GJW) constructed traversable wormholes in the context of the AdS/CFT correspondence \cite{Gao:2016bin}. 
  
  They began with an eternal AdS black hole, which contains an Einstein-Rosen bridge (wormhole) which is marginally non-traversable. This geometry is dual to two CFT's entangled in the thermofield double state \cite{Maldacena:2001kr}.  As we will review in the next section, they added a  coupling between the two CFTs. From the gravity perspective, this is a non-local coupling between the left and right asymptotic regions. 
This non-local coupling allows for negative null energy and makes  the wormhole traversable. 

The result of GJW provides a proof of existence for traversable wormholes in holography. Yet, the more fundamental question still remains to be answered: {\textit{what are the general rules for traversable wormholes?}} In this paper we take a step towards answering this question by analyzing the amount of information that can be sent through GJW-type wormholes.

First, we clarify some aspects of the GJW wormhole geometry. We calculate the time that the wormhole is open, defined as the maximum proper time separation between the past and future event  horizons, finding that this time is shorter than the Planck time.  While this result might suggest that the GJW wormhole cannot be trusted, we explain why it can still be analyzed within the semiclassical regime despite apparent Planckian features.\footnote{We thank Daniel Jafferis for discussions on this point.}

Next, we perform a bulk estimate of the amount of information that can be sent through the wormhole.  We show that for the original GJW construction, in which the two CFT's are coupled by a single operator, the amount of information we can transfer through the wormhole is proportional to the number of thermal cells 
\be
N \sim \left(r_{h} \over \ell\right)^{d-1} \,,
\label{boundeq}
\ee
in agreement with  \cite{Caceres:2018ehr}.
 Here, $r_{h}$ is the black hole radius, $\ell$ is the AdS radius and $d$ are the boundary spacetime dimensions. 
 
 The procedure to send this number of messages is to use modes with low angular momentum, such that the signal varies on scales somewhat longer than the AdS length scale in the transverse directions. The  message should be sent from the boundary long before the coupling between the two boundaries is turned on, so that it is very close to the horizon when it encounters the negative energy.
 
 In order to derive the bound (\ref{boundeq}), we impose a number of consistency conditions to remain in a controlled regime. In particular, following  \cite{Caceres:2018ehr}, we impose the `probe approximation': the message should backreact on the geometry by a small amount, so that the negative stress-energy tensor calculated in the absence of the signal is a good approximation. The probe approximation, in combination with our other conditions, allows us to do a well-controlled bulk analysis. However, as we discuss in more detail in the discussion section, it is not completely clear whether this condition must be imposed.

The capacity of the channel can be increased by including non-local couplings for a large number, $K$, of fields, as in \cite{Maldacena:2017axo}. In fact, many fields must contribute to the negative energy in order for the semiclassical description to be good.  In particular, in order to talk about a single metric sourced by the expectation value of the stress tensor, the fluctuations in the stress tensor should be small compared to the mean. We will see that meeting this condition requires a large number $K$ of coupled light fields. 

The opening of the wormhole increases linearly in $K$, and so does the amount of information we can transfer. However, a black hole has finite entropy, so there should be an upper bound on $K$. We show that $K \lesssim \ell^{d-1}/G_N$ is needed for a self-consistent bulk solution, where $G_N$ is the Newton constant. This bound can also be found by requiring that the UV cutoff of the theory is not lowered to the AdS scale. 

The final result is that the amount of transferable information is bounded by of order the entropy of the black hole $N \lesssim S_{BH}$, as expected. In order to send this amount of information,  we have to go beyond the s-channel and consider messages that are somewhat localized along the horizon. In particular we show that it is inefficient to localize messages on sub-AdS scales in the transverse directions, but it is possible if we couple many fields $K$. This would be needed for the comfortable journey of a cat through the wormhole envisioned by Maldacena, Stanford, and Yang  \cite{Maldacena:2017axo}.

We show that this result for the amount of information transfer is in accordance with CFT expectations coming from quantum teleportation. From the quantum theory perspective, the GJW protocol should be seen as analogous to quantum teleportation \cite{Maldacena:2017axo, Bak:2018txn}. Indeed, the thermofield double state is a specific pure, entangled state of two copies of the CFT. 
Roughly, each thermal cell of the left CFT, with size $\beta$, is entangled with the corresponding thermal cell in the right CFT. 

As in the standard qubit teleportation scenario, entanglement is not enough to send information from one copy to the other one: classical communication is also needed. Here, this is provided  by the $K$ couplings; these $K$ couplings play the role of approximately $K$ classical bits of information per thermal cell. Because the left-right coupling is local in space, it acts locally on the pairs of thermal cells. 
So at this rough level, the information transfer is simply standard quantum teleportation done on many qubits at once. 

Note one miracle that has to occur: although the thermal cells have a pairwise entanglement of $S_{cell}$, separating this into $S_{cell}$ EPR pairs would require solving a difficult problem at strong coupling. However, the $K$ couplings between left and right simply couple primary operators on each side. The miracle is that this crude coupling is sufficient for the delicate task of quantum teleporting a large amount of information, as discussed in more detail in \cite{Susskind:2017nto}.
This miracle is the same miracle that allows for the preparation of the TFD state from a simple Hamiltonian \cite{Maldacena:2018lmt, Cottrell:2018ash, Martyn:2018wli}.

The large value of $K$ that maximizes the information transfer is of order the entropy of an AdS size black hole, $K \sim \ell^{d-1}/G_N$ . For this value of $K$, the teleportation process uses up all of the quantum entanglement, destroying the black hole in the process.

Before continuing with our discussion, let us briefly comment on previous work related to traversable wormholes. It is by now well known that classical matter obeying the null energy condition, cannot support traversable wormholes -- see, for instance, \cite{Friedman:1993ty}. But this statement is no longer true if we include quantum corrections, leaving open the possibility that traversable wormholes are possible in the real world \cite{Maldacena:2018gjk}. Earlier results on how to build traversables wormholes using exotic matter or higher curvature theories of gravity include, among others, \cite{Visser:1989kh,Visser:1989kg, Poisson:1995sv, Barcelo:1999hq, Visser:2003yf, Bhawal:1992sz, Thibeault:2005ha,Arias:2010xg}. In the context of AdS/CFT, the fact that entanglement is not enough to build a traversable wormhole in AdS, but one needs an explicit coupling between the left and right asymptotic regions was already noted in \cite{Solodukhin:2005qy, Arias:2010xg}. After GJW, traversable wormholes were further explored for the case of AdS$_2$ in \cite{Maldacena:2017axo}, while recent attempts to construct eternal wormholes include \cite{Maldacena:2018lmt,Horowitz:2019hgb, Fu:2018oaq, Maldacena:2018gjk,Freivogel:2019lej}. The case of rotating wormholes in AdS was studied in \cite{Caceres:2018ehr}. Recently, the authors in \cite{Hirano:2019ugo}  found bounds on the information that can be transferred in the GJW wormhole. We will comment on the differences of both approaches in the discussion section.

The rest of the paper is organized as follows. In section \ref{secGJW} we give a quick review of the GJW construction and we explore several interesting facts that have not yet been pointed out in previous literature. In section \ref{secBounds} we bound the amount of information that can be transmitted through the wormhole in the bulk and check that it agrees with the boundary calculation. We conclude in section \ref{sec_conc} indicating some possible future directions.

\section{Gao-Jafferis-Wall wormhole} \label{secGJW}
In this section we review the traversable wormhole geometry constructed by Gao, Jafferis and Wall \cite{Gao:2016bin}. 
The main ingredient is a non-local coupling between the two asymptotic boundaries of a background BTZ black hole. This geometry contains an Einstein-Rosen bridge which is marginally non-traversable: photons falling along the horizon almost put in causal contact the two boundaries. The non-local coupling can generate a distribution of negative energy, which backreacts on the geometry such that these photons, if send early enough, can now fall from one boundary to the other. The wormhole becomes traversable.

The resulting geometry can be interpreted as the dual of the teleportation protocol \cite{Maldacena:2017axo}. The BTZ black hole is dual to two copies of a CFT entangled in a particular state, the thermofield double state (TFD)  \cite{Maldacena:2001kr}, which is defined by
\begin{eqnarray}
| TFD \rangle \equiv \frac{1}{\sqrt{Z}} \sum_n e^{-\beta E_n/2} |n\rangle_L \otimes |n \rangle_R \,,
\end{eqnarray}
where $|n\rangle_{L,R}$ are energy eigenstates of the left and right CFT's with energy $E_n$. This state is a pure entangled state from the perspective of the full system with the property that the reduced density matrix for each side is thermal with inverse temperature $\beta$. Details on how to create this state are given in \cite{Maldacena:2018lmt, Cottrell:2018ash, Martyn:2018wli}. The entanglement is the key resource for quantum teleportation, geometrically it \textit{builds} the connected wormhole geometry \cite{VanRaamsdonk:2010pw}. However, it is not enough, the exchange of classical information is also needed. The non-local coupling takes care of this second passage \cite{Maldacena:2017axo}, geometrically it makes the wormhole traversable.

We begin this section by recalling some basic properties of the unperturbed BTZ geometry.

\subsection{Unperturbed BTZ geometry}
The metric of the uncharged, non-rotating BTZ black hole is given by \cite{Banados:1992wn,Banados:1992gq},
\begin{equation}
\label{btzschw}
ds^2=-\frac{r^2-r_h^2}{\ell^2}dt^2+\frac{\ell^2}{r^2-r_h^2}dr^2+r^2d\phi^2,
\end{equation}
where $r_h$ is the horizon radius, $\ell$ is the radius of AdS and $\phi$ should be periodically identified $\phi \sim \phi + 2\pi$. The black hole mass and horizon radius are related by $r_h^2=8MG_N \ell^2$. The inverse temperature is given by $\beta= \frac{2\pi \ell^2}{r_h}$. $G_N$ is Newton's constant, which, in three dimensions, is related to the Planck length by $\ell_{P} = 8\pi G_N$. We use the convention that time is flowing upwards at the right boundary and downwards at the left one.
\begin{figure}[t]
	\centering
	\includegraphics[width=0.75\textwidth]{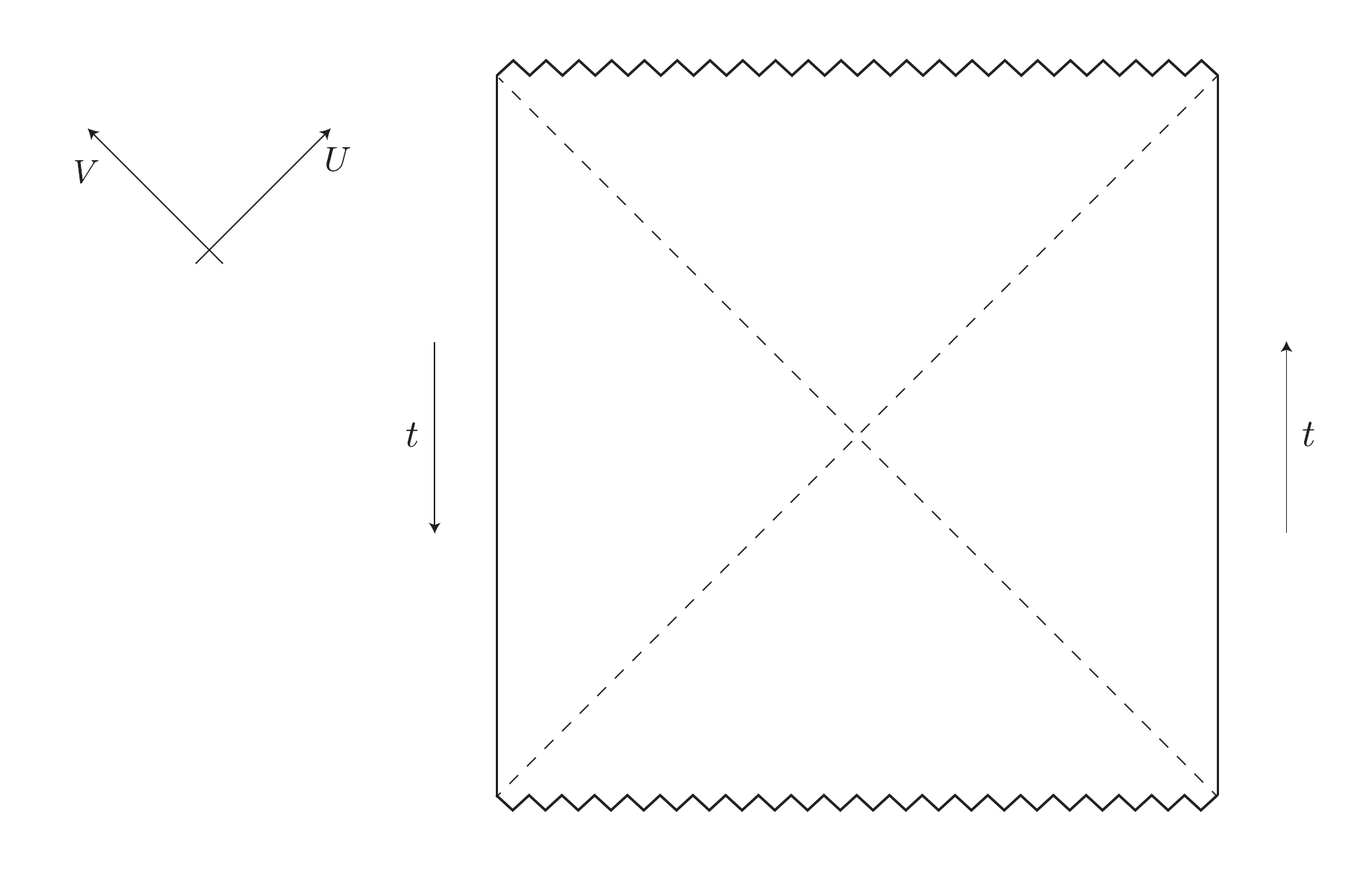}
	\caption{Penrose diagram for the BTZ black hole.}
	\label{fig:BTZPenrose}
\end{figure}

For our purposes, it will be convenient to work in Kruskal coordinates,
\begin{equation}
\label{krus}
\exp\!\left(2 \frac{r_h t}{\ell^2}\right)=-\frac{U}{V},~~~~\frac{r}{r_h}=\frac{1-UV}{1+UV},
\end{equation}
that cover the maximally extended two-sided geometry -- see Fig.\ \ref{fig:BTZPenrose} -- with the metric
\begin{equation}
\label{kruskalbtz}
ds^2=\frac{-4\ell^2dUdV+r_h^2\left(1-UV\right)^2d\phi^2}{\left(1+UV\right)^2},
\end{equation}
where $U>0$ and $V<0$ in the right wedge, $UV=-1$ at the boundaries and $UV=1$ at the two singularities. The two horizons correspond to $U=0$ and $V=0$ and, as can be seen from the figure, are on the verge of causally connecting the two boundaries. 

\subsection{Adding a non-local interaction}
\label{nonlocalint}
The main novelty of the GJW construction is a non-local interaction between the two boundaries, introduced through a small deformation of the original CFT Hamiltonian
\begin{eqnarray}
	\label{eq:HamilDeformation}
	\delta H = - \ell \int d\phi \, h(t,\phi) \, {\cal{O}}_L (-t,\phi) {\cal{O}}_R (t,\phi) \,. \label{pertS}
\end{eqnarray}
Here ${\cal{O}}_{L,R}$ are scalar primary operators with conformal weight $\Delta$ and $h (t,\phi)$ is the coupling constant. For the interaction to be relevant we need $\Delta<1$. In principle, the coupling constant could have some explicit dependence on $t$ or $\phi$, but we will restrict to a constant coupling $h$ turned on for some period of time
\begin{eqnarray}
h(t,\phi)=\begin{cases}
               h \left(\frac{2\pi}{\beta}\right)^{2-2\Delta} \quad &\text{if} \,\, t_0 \leq t \leq t_f \,,\\
              0 \quad &\text{otherwise} \,.
            \end{cases}
\end{eqnarray}
Note that, as in GJW, the factors of $\beta$ are chosen so that $h$ is dimensionless. We will consider the perturbative problem with $h \ll 1$.
As we explained above, a light ray falling along the horizon almost puts in causal contact the two boundaries. We want to show that the non-local coupling can be arranged in such a way that the backreacted null geodesic makes it from one boundary to the other. For simplicity we consider a radial geodesic, defined by $V=0$. 

First, we need to compute the expectation value of the stress-energy tensor along this geodesic. Since we will be interested in computing the shift in the $V$ direction, we only need to find $T_{UU}$.  The perturbation is spherically symmetric, hence $T_{UU}$, along $V=0$, can only depend on $U$. Once $T_{UU}$ is obtained, we compute the averaged null energy (ANE) by integrating it over the null ray, 
\begin{equation}
{\text{ANE}} (h,U_0,U_f) \equiv \frac{h}{\ell} {\cal{A}}(U_0,U_f) \equiv \int_{U_0}^{\infty} \langle T_{UU} \rangle (U) \, dU \,,\label{ANE}
\end{equation}
where $U_0$ and $U_f$ are the starting and ending times of the perturbation and, for later convenience, we have defined a dimensionless ANE, $\cal{A}$. Notice that in the last expression the dependance on the ending time of the perturbation $U_{f}$ is implicit in the definition of the expectatiation value of $T_{UU}$.

This will be our main diagnosis of the wormhole traversability. As we will see later, a negative ANE ``opens'' the wormhole by a magnitude proportional to the amount of averaged negative energy. So in order to quantify how traversable the GJW wormhole is, it is important to understand what are the optimal configurations and how much negative energy can we obtain from those. 

\begin{figure}
  \centering
  \includegraphics[width=0.8\textwidth]{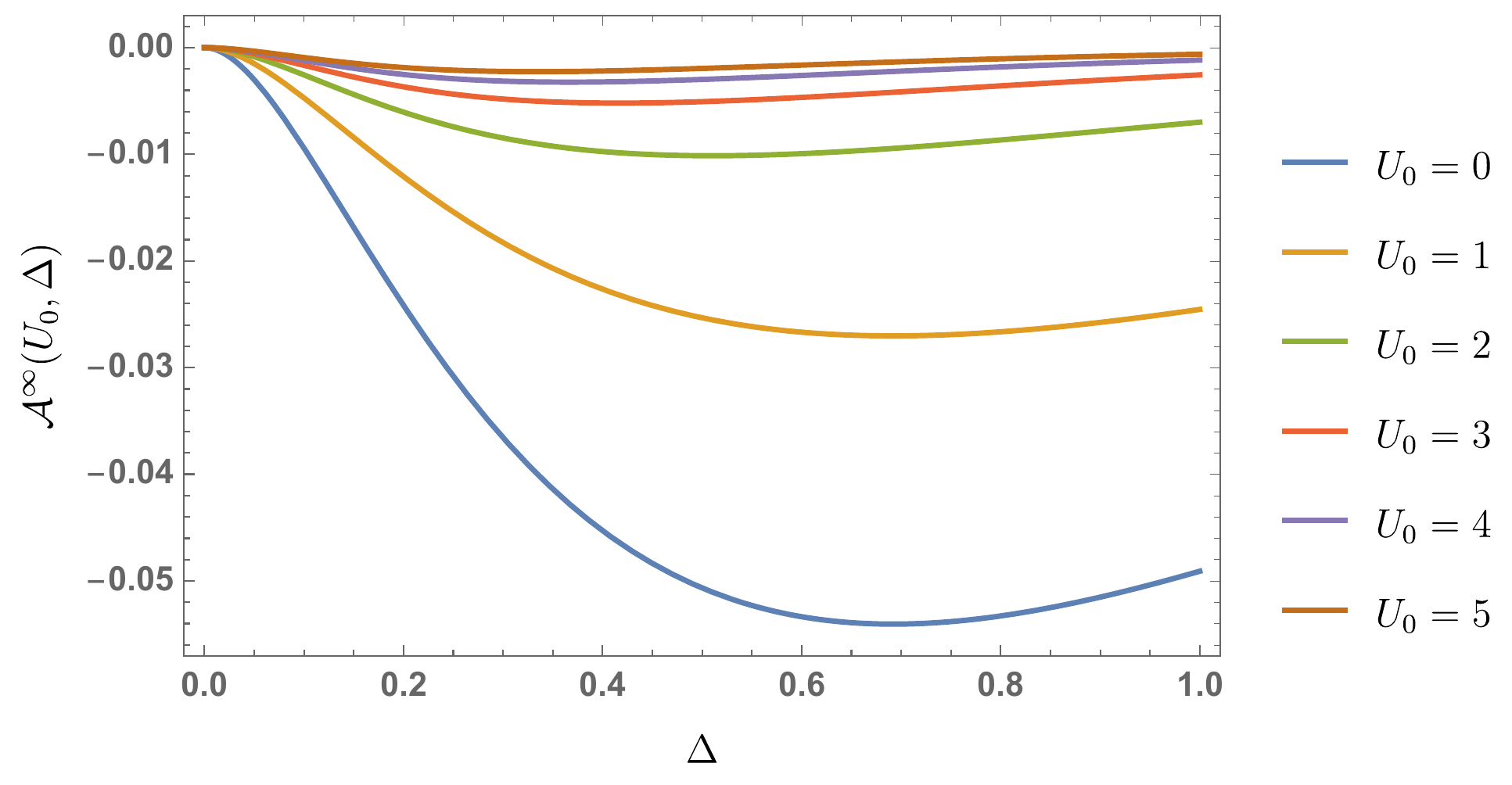}
  \caption{Dimensionless Averaged Null Energy as a function of $\Delta$. The curves correspond to different starting time $U_0$, while the end point is always $U_{f}=+\infty$. The earlier we turn on the interaction the more negative energy we can obtain (note that $U_0=0$ corresponds to boundary time $t=-\infty$ and $U_0=1$ to $t=0$).}
  \label{fig:longlived}
\end{figure}

We compute the stress-energy tensor by point-splitting, hence we need to find the modified bulk-to-bulk two-point function in the presence of $\delta H$ \eqref{pertS}. As usual in holography, the operators  ${\cal{O}}_{L,R}$ are dual to a bulk scalar field $\varphi$ with mass $m^2 \ell^2 = \Delta (\Delta - 2)$. Working on the right wedge it is possible to compute the modified propagator with
\begin{equation}
\begin{split}
G_h & \equiv  \langle \varphi^H_R(t,r,\phi)\varphi^H_R(t',r',\phi')\rangle_h =\\
& = \langle u^{-1}(t,t_0) \varphi^I_R(t,r,\phi)u(t,t_0)u^{-1}(t',t_0)\varphi^I_R(t',r,\phi)u(t',t_0)\rangle_h \,.
\end{split}
\end{equation} 
Here the subscript $R$ indicates that we are in the right wedge; the subscript $h$, that we are looking for the leading order correction in $h$; $H$ and $I$ indicate the Heisenberg and interaction fields respectively and $u(t,t_0)=\mathcal{T}\exp\left\{-i\int_{t_0}^{t}d\tilde{t}~\delta H(\tilde{t})\right\}$ is the evolution operator in the interaction picture. The result to leading order in $h$  is \cite{Gao:2016bin}
\begin{eqnarray}
G_h = 2 \sin \pi \Delta \int dt_1 h(t_1) K_\Delta(t'+t_1-i \beta/2) K_\Delta^r (t-t_1) +(t \leftrightarrow t') \,, \label{bulktobulk}
\end{eqnarray}
where $K_{\Delta}$ and $K_{\Delta}^{r}$ are the bulk-to-boundary and the retarded bulk-to-boundary propagators. Notice that we are omitting the $r,\phi$ dependence for simplicity. The propagators are  known analytically for the BTZ black hole\footnote{We suppress the sum over images in both propagators. When computing $G_{h}$ we can include one of these sums by extending the domain of integration over $\phi$ from $[0,2\pi]$ to the real line. We checked that the other sum gives contributions exponentially suppressed by $e^{-n\Delta r_{h}/\ell}$, where $n$ is the index that runs over images.} \cite{Ichinose:1994rg, Lifschytz:1993eb}
\begin{align}
K_{\Delta}\left(t,r,\phi\right)&=\left(\frac{r_h}{\ell^2}\right)^\Delta \frac{1}{2^{\Delta+1}\pi}\left(-\frac{\left(r^2-r_h^2\right)^{1/2}}{r_h}\cosh{\frac{r_h}{\ell^2}t}+\frac{r}{r_h}\cosh{\frac{r_h}{\ell}\phi}\right)^{-\Delta}\,,\\
	K_\Delta^r (t,r,\phi)& = |K_\Delta (t,r,\phi)| \theta{(t)} \theta{\left(\frac{\left(r^2-r_h^2\right)^{1/2}}{r_h}\cosh{\frac{r_h}{\ell^2}t}-\frac{r}{r_h}\cosh{\frac{r_h}{\ell}\phi} \right)}\,,
\end{align}
where $\theta$ is the Heaviside step function. We then need to transform equation \eqref{bulktobulk} into Kruskal coordinates and apply the point splitting formula to find the change in the expectation value of the stress tensor induced by the interaction at the horizon
\begin{eqnarray}
\langle T_{UU} \rangle (U) = \lim_{U'\rightarrow U} \partial_U \partial_{U'} G_h (U,U') \, .
\end{eqnarray}
For details on how to obtain explicit expressions for the stress tensor, range of validity and integrability properties, we refer the courageous reader to \cite{Gao:2016bin}. 
For sources which are turned on at time $t_{0}$ and left on forever, the stress-energy tensor is given by
\begin{equation}
\label{eq:negativeTUU}
\begin{split}
	\langle T_{UU} \rangle (U) =& -\frac{h}{\ell} \frac{2^{\frac{1}{2}-2 \Delta } \Delta  \sin (\pi  \Delta ) \Gamma (1-\Delta )}{\pi ^{3/2} \Gamma \left(\frac{3}{2}-\Delta \right)} \times \\
	&\times \lim_{U' \rightarrow U}  \partial_U \int_{U_0}^{U}dU_1 \frac{F_1 \left(\frac{1}{2};\frac{1}{2},\Delta+1;\frac{3}{2}-\Delta;\frac{U_1-U}{2U_1},\frac{U_1-U}{U_1(1+U'U_1}\right)}{U_1^{-\Delta+1/2}(U-U_1)^{\Delta-1/2}(1+U'U_1)^{\Delta+1}}\,,
\end{split}
\end{equation}
where $F_1$ is the Appell hypergeometric function. Now, to compute the ANE, we need to further integrate the above expression along $U$. Surprisingly, this can be done analytically \cite{Gao:2016bin}  
\begin{eqnarray}
\label{eq:ANECInfinite}
\begin{split}
	{\cal{A}}^\infty (U_0,\Delta) & \equiv   \frac{\ell}{h} \text{ANE} (h, U_0, \infty) =  \\
& =- \frac{\Gamma (2 \Delta+1)^2 }{2^{4 \Delta} (2 \Delta +1) \Gamma (\Delta)^2 \Gamma (\Delta+1)^2} \frac{_2F_1\left(\Delta+\frac{1}{2},\frac{1}{2}-\Delta;\Delta+\frac{3}{2};\frac{1}{1+U_0^2}\right)}{\left(1+U_0^2\right)^{\Delta+\frac{1}{2}}} \,,
\end{split}
\end{eqnarray}
where now the $_2F_1$ is the ordinary hypergeometric function.
It is instructive to plot the ANE to see how it depends on the different parameters involved. This is done in Fig.\ \ref{fig:longlived}, where we plot the ANE as a function of $\Delta$ for different starting times $U_0$. 
As expected, the sooner we turn on the coupling the larger amount of negative energy we can get. The curve with $U_0=0$, \textit{i.e.}\ $t_{0}=-\infty$, gives an upper bound on the amount of negative energy we can get with this type of sources: $|{\cal{A}}^\infty (U_0,\Delta)| \lesssim 10^{-1}$.
One might worry that if the source is turned on for such a long time we should take into consideration the backreaction of the negative energy on the geometry. However, one can check that the gravitational perturbation is small everywhere and so the linear order computation can be trusted. 

The analytical expression \eqref{eq:ANECInfinite} found by GJW is a remarkable result; however, it is somewhat impractical to deal with hypergeometric function. In particular in the next section we will need to quantify the backreaction of a message on the quantity of negative energy and it would be helpful to have at disposal a simpler expression for the ANE. In the following we provide such a simple analytic expression, valid for the case of instantaneous sources 
\begin{eqnarray}
	h^{inst}(t,\phi) = h \, \left(\frac{2\pi}{\beta}\right)^{2-2\Delta} \delta \left(\frac{2\pi}{\beta}(t-t_0)\right) \,. \label{deltah}
\end{eqnarray}
This can be found by manipulating equation \eqref{eq:ANECInfinite}. First, we find an expression of the ANE for \textit{smeared} interactions. This is a more physical scenario in which we turn on the sources only for a finite amount of time $\Delta U = U_f - U_0$. 
As before, we define a dimensionless ANE,
\begin{eqnarray}
{\cal{A}}^s (U_0, U_f, \Delta) & \equiv &   \frac{\ell}{h} \text{ANE} (h, U_0, U_f) \,,
\end{eqnarray}
where the subscript $s$ stands for smeared. The ANE involves integrating over a whole null ray the stress-energy tensor, that by itself is an integral over the sources. Schematically, we can write this as
\begin{eqnarray}
{\cal{A}}^s (U_0, U_f, \Delta) = \int_{U_0}^\infty dU \langle T_{UU}\rangle(U) \equiv \int_{U_0}^\infty dU \int_{U_0}^{U_f} dU_1 \, \tau(U,U_1) \,
\end{eqnarray}
where we have defined a function $\tau(U,U_1)$ whose integral is the stress energy tensor. Notice that this quantity is allowed to have some discontinuities at the positions where the sources are turned on/off. We can rewrite ${\cal{A}}^s$ in terms of ${\cal{A}}^\infty$ as follows
\begin{equation*}
\begin{split}
{\cal{A}}^s (U_0, U_f, \Delta)=&\int_{U_0}^\infty dU \left[ \int_{U_0}^{\infty} dU_1 \, \tau(U,U_1) - \int_{U_f}^{\infty} dU_1 \, \tau(U,U_1) \right] =\\
= & \left( \int_{U_0}^\infty dU \int_{U_0}^{\infty} dU_1 -\cancel{ \int_{U_0}^{U_f} dU \int_{U_f}^{\infty} dU_1} - \int_{U_f}^{\infty} dU  \int_{U_f}^{\infty} dU_1 \right) \tau(U,U_1)=\\
=&\; {\cal{A}}^\infty (U_0, \Delta) - {\cal{A}}^\infty (U_f, \Delta) \,.
\end{split}
\end{equation*}
The integral in the second line vanishes because it adds up the energy generated along the null geodesic for $U<U_f$ by a source turned on only at $U_f$, \textit{i.e.}\ the support of the first integral lies outside the lightcone of the second.
Finally, we take the limit in which the coupling is turned on only for an instant of time and find a remarkably simple analytic expression. 
The dimensionless ANE in this case is given by\footnote{The extra $U_0$ on the RHS comes from the fact that the source is a $\delta$-function in time while for this limit we are taking a $\delta$ in the $U$-coordinate.}
\begin{eqnarray}
{\cal{A}}^{inst} (U_0, \Delta) = \lim_{U_f \rightarrow U_0} U_0 \frac{{\cal{A}}^{s}(U_0,U_f,\Delta)}{U_f-U_0} = - U_0 \, \partial_{U_0}{\cal{A}}^{\infty} (U_0,\Delta) \,.
\end{eqnarray}
It is straightforward to evaluate this expression: all the dependence on $U_0$, that before was encoded in the hypergeometric functions, now becomes simply
\begin{eqnarray}
{\cal{A}}^{inst} (U_0, \Delta) = -\frac{ \Gamma \left(\Delta+\frac{1}{2}\right)^2}{\pi  \Gamma (\Delta)^2} \left(\frac{U_0}{1+U_0^2}\right)^{2 \Delta+1} \,. \label{Ainst}
\end{eqnarray}

Notice that this expression might receive large correction from higher order terms in $h$, which should be cured by introducing a small smearing. Nonetheless, it provides a simple and helpful approximation to the smeared case. Also note that in this way of deriving equation \eqref{Ainst}, we did not need to directly integrate the stress-energy tensor for the instantaneous source, which was computed in \cite{Dora}, and has an apparent non-integrable divergence for $\Delta>1/2$.

\begin{figure}[h!]
        \centering
        \begin{subfigure}[t]{0.75\textwidth}
        		\centering
                \includegraphics[width=\textwidth]{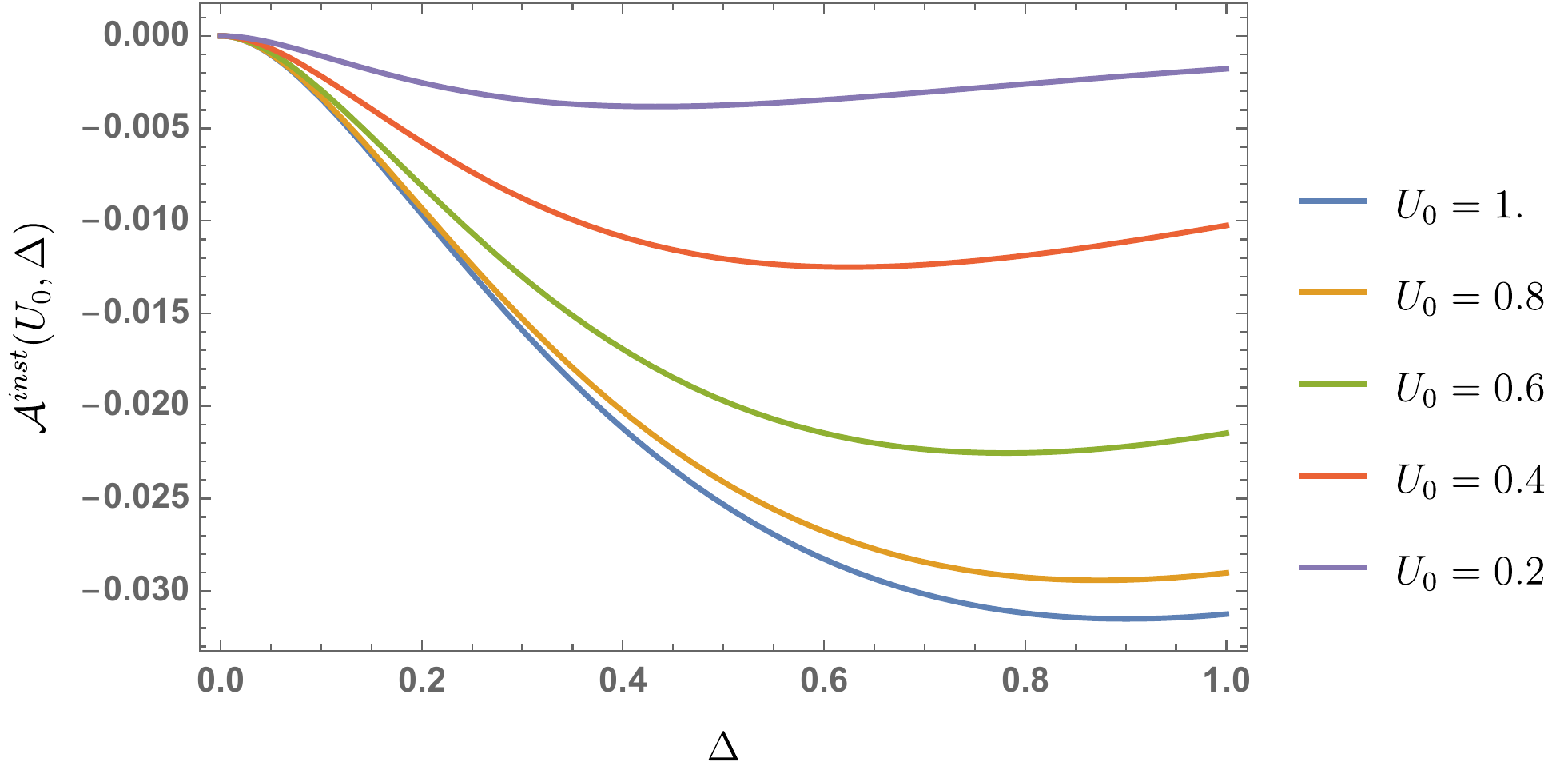} 
                \caption{}
        \end{subfigure}
        \par\bigskip
         \begin{subfigure}[t]{0.75\textwidth}
        		\centering
                \includegraphics[width= \textwidth]{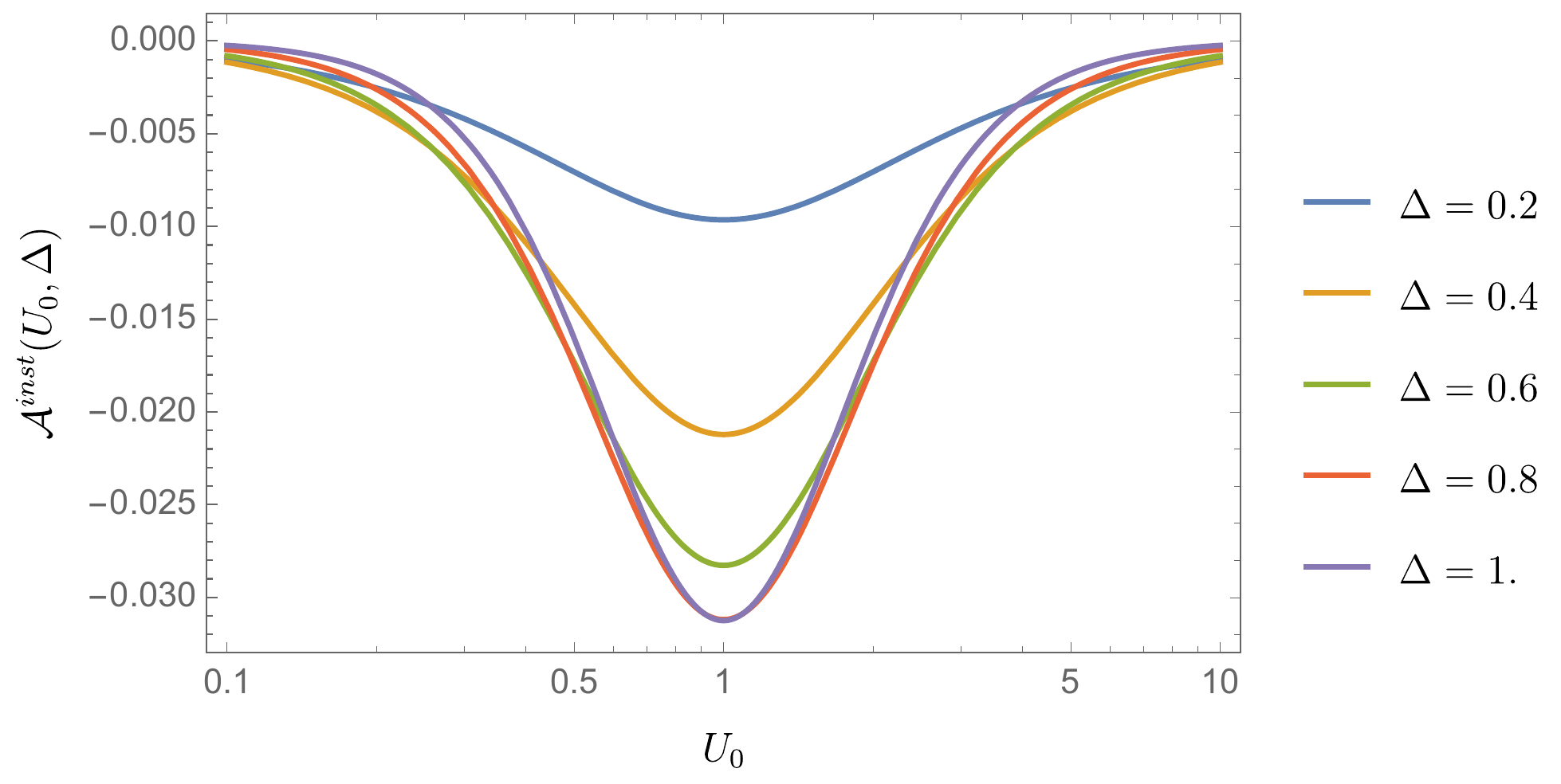} 
                \caption{}
        \end{subfigure}
                 \caption{Averaged Null Energy for the case of instantaneous sources. In (a) we plot the ANE as a function of $\Delta$ for different $U_0$. The largest amount of ANE (in absolute value) is given by the curve with $U_0=1$. This corresponds to $t_{L}=t_{R}=0$. Note that for this choice of times, the wormhole is shortest; therefore, it is not surprising that the effect of the non-local coupling is the largest. In (b), we plot the ANE as a function of $U_0$ for different $\Delta$. The scale on the $U_{0}$ axis is logarithmic which makes evident that each curve has a minimum at $U_0=1$ and is symmetric under $U_0 \rightarrow U_0^{-1}$.}
\label{fig:inst}
\end{figure}

Plots of ${\cal{A}}^{inst}$ are shown in Fig.\ \ref{fig:inst}. Note that this expression has a few interesting properties. First, ${\cal{A}}^{inst}$ is symmetric under $U_0 \rightarrow U_0^{-1}$, which makes the time $U_0=1$ somewhat special. In fact, it is easy to show analytically that $U_0=1$ is a minimum of this function. It is also possible to note that the optimal weight, for $U_{0}=1$, is $\Delta \approx 0.9$. Finally, we observe that for $\delta$-function source $|{\cal{A}}^{inst} (U_0,\Delta)| \lesssim 10^{-2}$. This means that turning on the non-local coupling for only an instant of time, at $U_{0}=1$, reduces the amount of negative energy by only an order of magnitude, as compared to ${\cal{A}}^\infty$. 

Independently of the details of the interaction, we see that the averaged null energy of the Gao-Jafferis-Wall protocol is bounded, in absolute value, by $\lvert h\rvert/\ell$ times an order-one number. More precisely, we have
\begin{eqnarray}
\label{eq:ANE}
|\text{ANE}(h,U_0,U_f)| = \frac{|h|}{\ell} |{\cal{A}}| \lesssim \frac{|h|}{\ell} 10^{-1} \,.
\end{eqnarray}
Given that we are working perturbatively in $h$, we conclude that the ANE is generically very small in AdS units. This fact is worrisome because, as we show below, the amount of available negative energy determines the size of the wormhole opening. 

Before continuing with our analysis let us clarify a point concerning the reference frame we have considered so far. We have worked in a frame in which $t_{L}=-t_{R}$, that we will call the \textit{rest} frame. The BTZ geometry is invariant under boosts, $\{U\rightarrow \lambda U, V\rightarrow \lambda^{-1} V\}$, which act on the asymptotic boundaries as time translations $t_{L,R}\rightarrow t_{L,R}+\delta t$. Therefore, we can consider more general reference frames in which $t_{L}\neq-t_{R}$. We will call these, \textit{boosted} frames. This is just a change of coordinates, so the bound on information transfer we are looking for should be independent of $\lambda$. However, the integrated null energy, and hence, the dimensionless coefficient $\cal{A}$  is not invariant under these boosts. 
In fact, an expression for ${\cal{A}}$ in generally boosted frames can be easily found, and in Schwarzschild-like coordinates, is given by
\begin{eqnarray}
\label{eq:instANE}
{\cal{A}}_{\rm boosted}^{inst} (t_{R}, t_{L}, \Delta) = - \exp \left({-\frac{\pi}{\beta}(t_{L}+t_{R})}\right)\frac{ \Gamma \left(\Delta+\frac{1}{2}\right)^2}{\pi  \Gamma (\Delta)^2} \left[\frac{1}{2}\cosh\left(\frac{\pi}{\beta} (t_{R}-t_{L})\right)\right]^{-2 \Delta-1} \,.
\end{eqnarray}
The exponential on the r.h.s.\ corresponds to the boost factor $\lambda$ we discussed above. Indeed, we see that in the boosted frames we can get much more negative energy than in the rest frame. In the next section we will explain why this, as expected, cannot enhance the amount of information we can transfer across the wormhole. In fact, for simplicity, we will keep working in the rest frame, eventually deriving a coordinate-independent expression for the opening of the wormhole -- see equation \eqref{delta_tau}.

\subsection{Wormhole opening}
We can relate the ANE to the opening size of the wormhole through the linearized Einstein equations. Let $g_{\mu\nu} = g_{\mu\nu}^{BTZ} + h_{\mu\nu}$, be the metric of the perturbed BTZ geometry, then in Kruskal coordinates, to leading order in $h_{\mu\nu}$ and at the horizon $V=0$, the linearized Einstein equation reads \cite{Gao:2016bin}
\begin{eqnarray}
\frac{1}{2} \left(\frac{U h_{UU}'(U)+2 h_{UU}(U)}{\ell^2}-\frac{h_{\phi\phi}''(U)}{r_h^2}\right) = 8\pi G_N \langle T_{UU} \rangle \,,
\end{eqnarray}
where the primes denote derivative with respect to $U$. In fact, in order to use this equation, where the classical metric is sourced by the expectation value of the stress tensor, the fluctuations in the stress tensor should be small compared to its mean. As we discuss in more detail in section \ref{section_swave}, this semiclassical condition is only met if a large number $K$ of fields contributes to the negative energy. We do not include the factors of $K$ for now, since we are reviewing the original construction, but will include them in our later analysis since they are essential for remaining in the semiclassical regime.

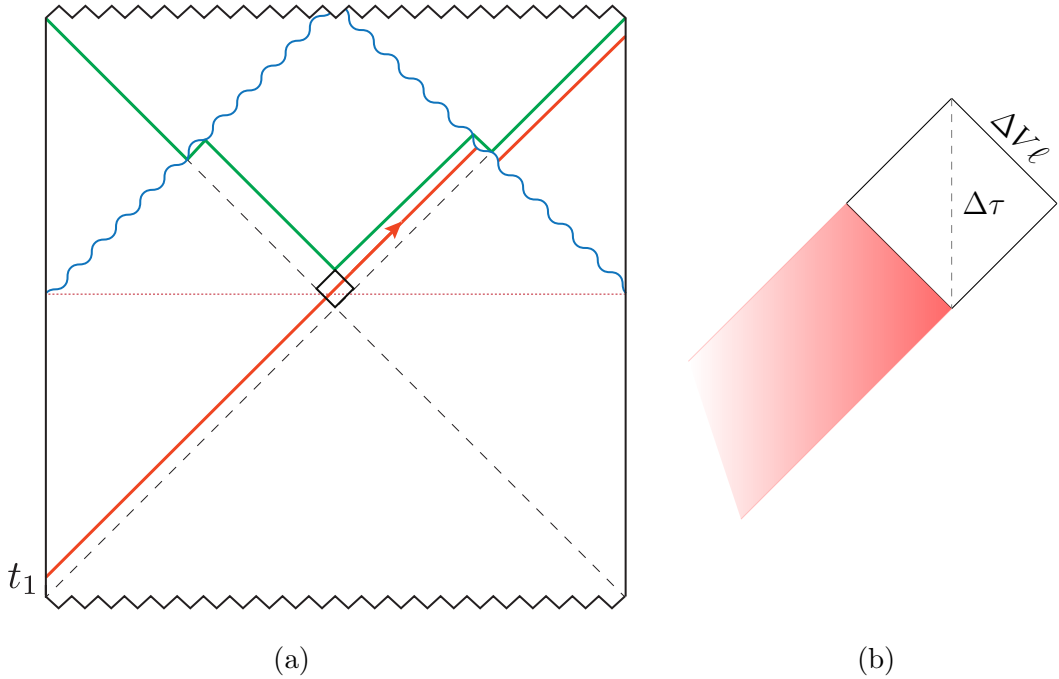
\begin{figure}
\centering
\begin{subfigure}[t]{.5\textwidth}
  \centering
  \includegraphics[width=1.1\linewidth]{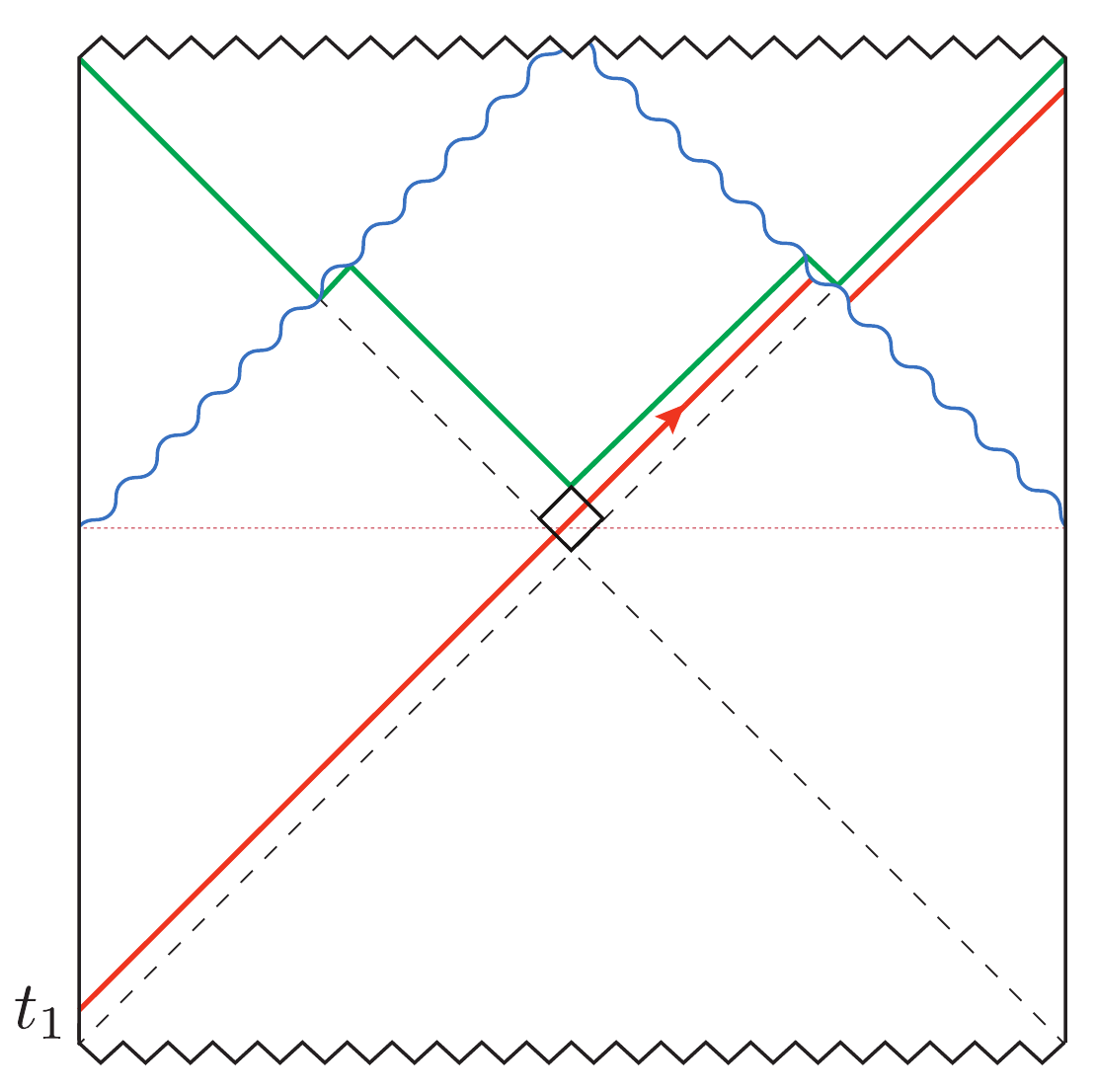}
  \caption{}
  \label{fig:sub1}
\end{subfigure}%
\begin{subfigure}[t]{.5\textwidth}
  \centering
  \begin{tikzpicture}[scale=0.70]
\draw[black]  (0,2)--(2,0);
\draw[black]  (2,0)--(4,2);
\draw[black]  (4,2)-- node [midway, above, sloped] {$\Delta V \ell$}(2,4);
\draw[black]  (2,4)--(0,2);
\draw[gray,dashed]  (2,4)--(2,0);
\node[right] at (2,2){$\Delta \tau$};
\draw[red!20] (0,2)--(-3,-1);
\draw[red!20] (2,0)--(-2,-4);
===\fill[red!60, path fading=west](0,2)--(-3,-1)--(-2,-4)--(2,0)--(0,2);
\draw[white]  (2,-6)--(2,-4);
\end{tikzpicture}
  \caption{}
  \label{fig:sub2}
\end{subfigure}
\caption{(a) The blue wavy lines represent the negative energy. The green line shows how the future horizon recedes due to the backreaction of the negative energy on the geometry. As the future horizon moves up, a diamond region is revealed in the middle of the diagram. The past horizon remains unaffected. The red line is the positive energy signal that we send through the wormhole. (b) Here, we have zoomed into the diamond region. The side of the diamond is equal to $\Delta V\ell$. $\Delta \tau$ is the amount of proper time that the wormhole remains open. The red region represents a signal that passes through the wormhole throat. }
\label{fig:test}
\end{figure}

 Integrating this equation, the total derivative terms do not contribute if the perturbation decays sufficiently fast at $U=\pm \infty$, which is the case unless the perturbation is turned on forever. We are left with
\begin{eqnarray}
\frac{1}{2 \ell^2} \int dU h_{UU} (U) = 8\pi G_N \int dU \langle T_{UU} \rangle \,.
\end{eqnarray}
A null ray traveling close to the horizon will suffer a shift in its $V-$coordinate, see Fig. \ref{fig:sub1}, due to the perturbation given by

\begin{eqnarray}
\Delta V = \frac{1}{4\ell^2} \int dU h_{UU} \,.
\end{eqnarray}
Combining both equations together we obtain
\begin{eqnarray}
\Delta V = 4 \pi G_N \int dU \langle T_{UU} \rangle = 4 \pi \frac{G_N h}{\ell} {\cal{A}} (U_0, U_f) \,
\end{eqnarray}
where the dimensionless ANE, $\cal{A}$, is defined in \eqref{ANE}. If $\Delta V <0$, a null ray traveling close to the horizon and starting on one side of the black hole will end up traversing the wormhole and appear on the other side. Given the analysis in the previous subsection, it is clear that the non-local interaction can make $\Delta V$ negative by a proper choice of the sign of $h$.  However, as explained above, ${\cal{A}}$ is frame dependent and so is $\Delta V$. We can quantify the opening of the wormhole in a coordinate independent way by computing the proper time that the wormhole remains open. 
To do this, we zoom into the diamond region that appears between the future and the past horizon due to the backreaction of the negative energy, see Fig.\ {\ref{fig:sub2}}. Near the horizon, the metric is approximately
$
ds^2\approx-4\ell^2 dU dV.
$
Consequently the proper time that separates the lower and upper vertices of the diamond region is
\begin{equation}
\Delta \tau \approx 2 \ell \sqrt{\Delta V\Delta U}.
\end{equation}
In the rest frame the coupling is symmetric under $L\leftrightarrow R$, hence $\Delta V=\Delta U$ and the above relation reduces to $\Delta \tau \approx 2 \ell \Delta V$. Combining everything, we have an upper bound on the proper time between the past and future event horizons,
\be
\Delta \tau \approx 8 \pi G_N h {\cal{A}} \lesssim G_N \label{delta_tau} \,,
\ee
where ${\cal{A}}$ is the one computed in the rest frame. In the boosted frames, the extra contributions coming from $\Delta U$ and $\Delta V$ will cancel perfectly, leaving the expression in the rest frame.
We conclude that the time window that the wormhole remains open is indeed Planckian, independent of the chosen frame. 

Since the time for which the  wormhole is open is so small, one might worry that quantum gravity corrections are important and cannot be neglected. This is not the case. The diamond is just a - small - piece of the BTZ geometry, the invariant curvature is given by $\ell^{-2}$ and is well separated from the Planck scale. While passing through the wormhole a signal would just feel like traveling through empty flat spacetime. Nonetheless, we still need to make sure that the signal is localized to a Planck sized box to be certain that it will make it through the opening. This sounds like a difficult, even dangerous, task. In this case we don't need to worry about this issue because the mouth of the wormhole is located close to the horizon of a black hole. The gravitational blueshift makes sure that an ordinary message at infinity is boosted enough by the time it reaches the mouth of the wormhole to fit in such a Planck sized box. We just need to send the message from the boundary early enough. The same gravitational effect guarantees that we don't need to fine-tune the moment we send the message from the boundary up to a Planck-time precision, because an asymptotic observer sees the window open for an exponentially longer time. We conclude that, despite the smallness of the opening, it is kinematically possible to send a message through the wormhole.\footnote{We thank Daniel Jafferis for discussions on this point.}

\section{Bound on information transfer} \label{secBounds}
In the previous section we have revised the construction of GJW. The non-local coupling between the two asymptotic boundaries is enough to open the wormhole for only a Planck-sized window of time. Nonetheless, we have argued that thanks to redshift this by itself is not an obstacle for a message traversing the wormhole. However, so far we have neglected the backreaction of the message on the geometry, staying in the so called \textit{probe approximation}. Given that the message needs to be highly boosted to make it through the wormhole, one might worry that this is not a valid approximation. In fact the signal might destroy the original traversable wormhole setup altogether. In this section we check that this is not the case. It is possible to send a large amount of information, while keeping under control the backreaction on the geometry. 

For convenience we now summarize the main results that will be proven throughout the section. We begin by finding a simple condition on the total momentum that we can send through the wormhole before the probe approximation becomes unreliable. Because the signal is highly boosted by the time it reaches the horizon, it can be approximated by a shock wave and its stress energy tensor (in light-cone coordinates) is given by \cite{Shenker:2014cwa, Dray:1985yt}
\begin{equation}
T_{VV}=\frac{p_{V}}{r_{h}}\delta(V) \,, \label{tvv}
\end{equation}
where $p_V$ is the {\textit{total}} momentum of the message. In the next subsection we show that the probe approximation is valid as long as\footnote{Note that this statement is coordinate dependent. We will also provide an equivalent statement in terms of the center-of-mass energy collision in equation (\ref{probewiths}).}
\begin{equation}
\label{eq:probeApp}
\boxed{\text{probe approximation:} \,\,\, \frac{G_N \, p_V}{r_h} \ll 1\,.}
\end{equation} 

This, combined with the requirement that each message we send is boosted enough to fit through the wormhole opening, is enough to constraint the amount of information we can transfer. We can estimate how much one signal needs to be boosted using the {\textit{uncertainty principle}} \cite{Maldacena:2017axo}. 

First, we consider messages that are completely spread across the horizon, \textit{i.e.}\ s-waves.  We show that the amount of transferred information can be made large by increasing the radius of the black hole; however, it is always much smaller than the entropy of the black hole. To increase the number of bits that can be sent through the wormhole, for fixed $r_{h}$, we follow the approach of \cite{Maldacena:2017axo} and couple a large number $K$ of fields. Combining the probe approximation with the uncertainty principle we obtain that the number $N$ of bits that can go through the wormhole is given by
\begin{eqnarray}
N \lesssim \frac{r_h}{\ell} K \,.
\end{eqnarray}

However, $K$ cannot grow arbitrarily large. Treating the negative energy as a negative shock we find that the maximum number of coupled fields beyond which our construction becomes unreliable is 
\begin{equation}
\boxed{\text{species bound:} \,\,\, K \lesssim \frac{\ell}{G_N}\,.} \label{speciesbound}
\end{equation} 
We can reproduce this bound also by imposing that the renormalized UV cutoff, see \cite{Dvali:2007wp,Kaloper:2015jcz}, is above the AdS scale. 
Combining both results, we obtain the final bound on the information that can be sent through the wormhole,
\begin{eqnarray}
N \lesssim \frac{r_h}{G_N} \approx S_{BH} \,.
\end{eqnarray}
We check that this bound is consistent with the one we obtain by considering the boundary theory. Finally we consider signals that are localized in the transverse direction. We show that it is possible to localize messages on sub-AdS scales if we couple $K\gg1$ fields. In the rest of the section we give details on how to derive these bounds.

Notice that the breakdown of the probe approximation does not necessarily imply that the wormhole closes. It only means that the GJW computation is not reliable anymore. One might wonder if by fully taking into account the backreaction, we might increase the amount of transferable information. The analysis of \cite{Caceres:2018ehr} suggests that this is not the case; however, we believe that this issue has not been settled yet. We will comment further on this in the discussion section. 

\subsection{S-wave channel} \label{section_swave}
\begin{figure}
	\begin{center}
	\includegraphics[width=0.5\textwidth]{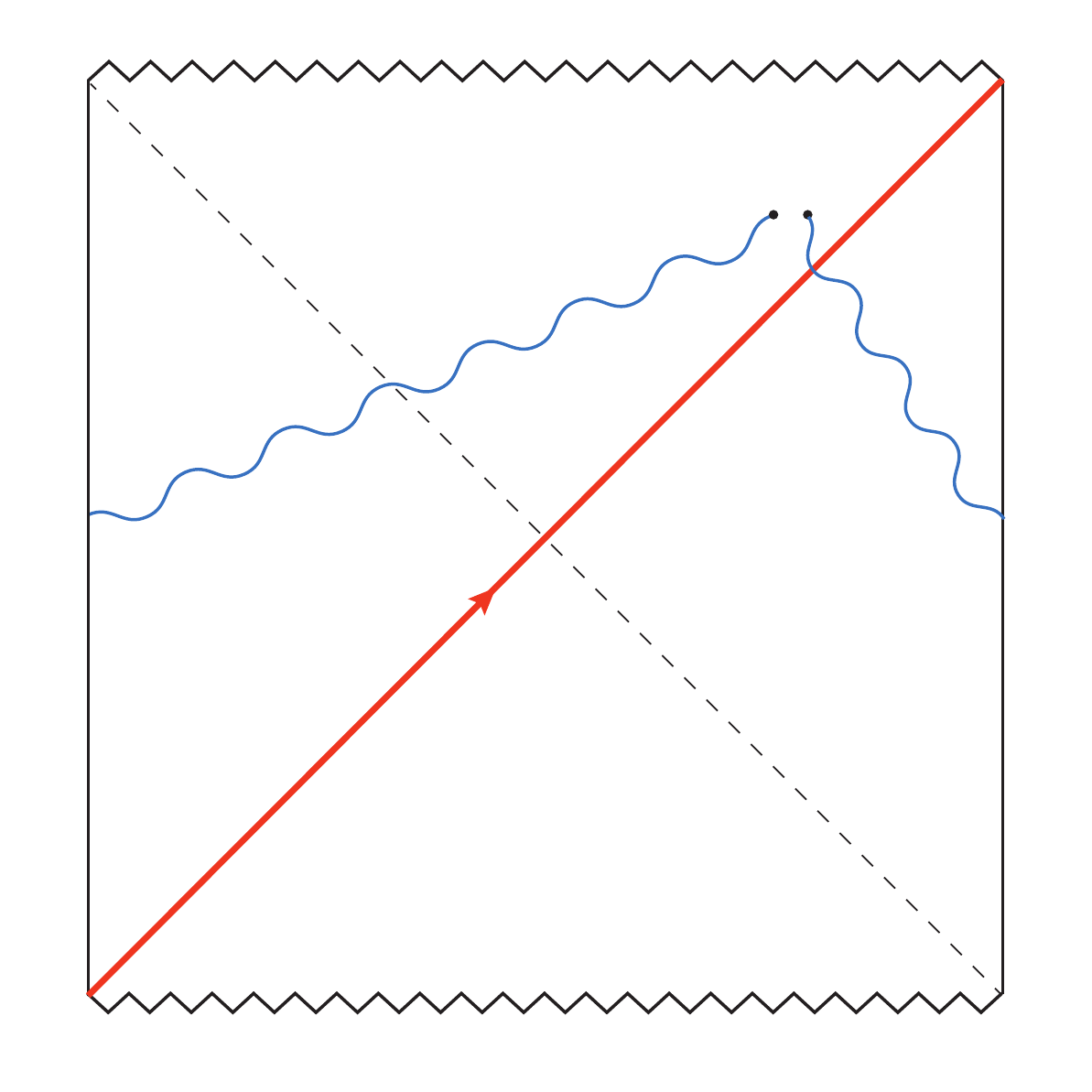}
	\begin{picture}(0,0)
	\put(-150,155){{\small $K_{\Delta}$}}
	\put(-45,155){{\small $K_{\Delta}^{r}$}}
	\put(-78,180){{\small $x$}}
	\put(-58,180){{\small $x'$}}
	\put(-221,115){{\small $t_{L}$}}
	\put(-18,115){{\small $t_{R}$}}
  	\end{picture}
	\caption{Schematically the ANE is given by the product of a boundary-to-bulk propagator coming from the left CFT and a retarded boundary-to-bulk propagator coming from the right one. Only the latter crosses the positive energy shock and undergoes a time delay.}
	\label{fig:shockANE}
	\end{center}
\end{figure}
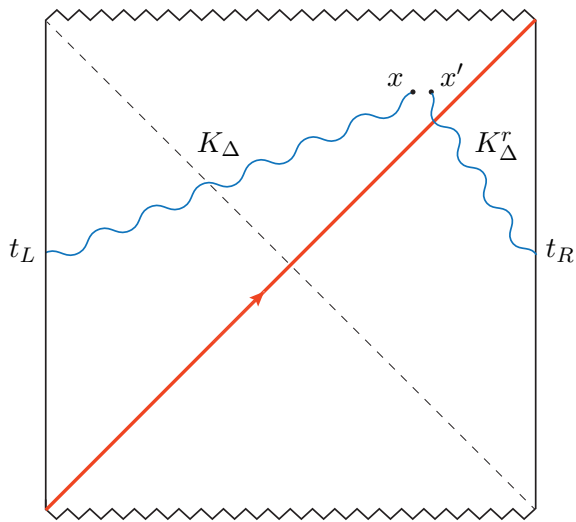

We want to bound the amount of information we can send through the wormhole. As we explained above, to do this we first need to understand how far we can trust the probe approximation. To begin with, we consider spherically symmetric messages. To estimate the effect of one such message on the amount of negative energy generated by the non-local coupling, we approximate the message as a positive energy shock, propagating along the horizon $V=0$.  At linear order we don't need to worry about the backreaction of the negative energy shock, the geometry is then simply given by \cite{Shenker:2013pqa}
\begin{equation}
\begin{split}
	ds^{2}=&\frac{-4\ell^{2}dUdV+r_{h}^{2}(1- (U+\Delta U\theta(V)) V)^{2} d\phi^2}{(1+(U+\Delta U\theta(V))V)^{2}} \,, \\
	=&\frac{-4\ell^{2}d\tilde{U}dV+4\ell^{2}\Delta U \delta(V)dV^{2}+r_{h}^{2}(1-\tilde{U}V)^{2}d\phi^2}{(1+\tilde{U}V)^{2}}\,,
\end{split}
\end{equation}
where in the second line we have used the discontinuous coordinate $\tilde{U}=U+\Delta U\theta(V)$. To compute the negative energy we need to know the propagator for the scalar field in this shockwave geometry. Away from $V=0$ this is simply given by the BTZ propagator as the geometry is the one of the BTZ black hole. However, the shockwave induces a discontinuity across $V=0$, one can check that it is enough to use the usual BTZ propagator but using the discontinuous coordinates,
\begin{equation}
	K^{\text{shock}}(U,V)=K^{\text{BTZ}}(\tilde{U},V)\,,
\end{equation}
where schematically $K$ is a BTZ propagator. When computing the ANE we evaluate two BTZ boundary-to-bulk propagators. But note that only the one coming from the right boundary crosses the positive energy shock and undergoes a time delay. See Fig.\ \ref{fig:shockANE}.
The effect of this delay is equivalent to shifting the insertion time of $O_{R}$ in \eqref{eq:HamilDeformation} by a quantity
\begin{equation}
	\Delta t \approx \beta \frac{\Delta U}{U_{0}}\,, 
\end{equation}
where we have assumed that the shift is small. The rest of the ANE computation of GJW follows unchanged. The probe approximation is valid as long as the effect of the message on the geometry can be neglected, \textit{i.e.}\ as long as $\Delta U\ll1$. The time delay is related to the stress energy tensor generated by the message by \cite{Dray:1985yt, Shenker:2013pqa}
\begin{equation}
	\label{eq:timeDelayT}
	T_{VV}=\frac{\Delta U}{G_{N}}\delta(V)\,.
\end{equation}
Comparing with \eqref{tvv} we can relate the time delay to the total momentum carried by the message 
\begin{equation}
	\label{eq:probeAppShocks}
	\Delta U=G_{N}\frac{p_{V}}{r_{h}}\ll1\,.
\end{equation}
We will see below that this momentum needs to be large enough such that the message can fit through the wormhole. Notice that the momenta in Kruskal coordinates are dimensionless since they are the conjugate variables to the dimensionless $U,V$-coordinates. We can check that this condition is enough to ensure that the negative energy is almost preserved by using our simple analytical expression \eqref{eq:instANE}, valid for the case of instantaneous sources. For simplicity we consider the optimal case in which $t_{R}-t_{L}=0$. We can trust the probe approximation if 
\begin{equation}
\label{eq:probeApp2}
	\left\vert\frac{\mathcal{A}(0,\Delta t)-\mathcal{A}(0,0)}{\mathcal{A}(0,0)}\right\vert\ll1\,.
\end{equation}
It is easy to see that this condition reduces to \eqref{eq:probeAppShocks}.

It is also possible to express the constraint above in terms of coordinate-independent quantities. We approximate the interaction between the negative energy density and the signal as a collision between particles and assume that the scattering is dominated by gravitational interaction. Along the horizon of a BTZ black hole, gravitational interaction decays exponentially outside a region of size $\ell$, see equation \eqref{eq:greenFunction}. 
This is expected since this region is a thermal cell on the horizon, \textit{i.e.}\ it corresponds to a region of size $\beta$ on the boundary. Therefore, we can split the collision in independent events, one per thermal cell. The probe approximation translates to the statement that the amplitude associated to each of these collision events should be small
\begin{eqnarray}
	G_N s_{cell} \ell \ll 1\,. \label{probewiths}
\end{eqnarray}
Here $s_{cell}=p_{V}^{cell}q_{U}^{cell}/\ell^{2}$ is the center-of-mass energy squared of one of these collisions. In the rest frame, the momentum of the negative energy, per thermal cell, is given by $q_{U}^{cell}\approx 1$, $p_{V}^{cell}$ is simply given by $p_{V}$ divided by the number of thermal cells, $r_{h}/\ell$. It is easy to see that with these identifications, equation \eqref{probewiths} reproduces \eqref{eq:probeApp}. 

The probe approximation provides an upper bound on the momentum a particle traversing the wormhole can have. As pointed out in the previous section, this particle needs to be highly boosted to fit through the wormhole, so the momentum cannot be arbitrarily small. We can estimate the minimum required momentum using the uncertainty principle\footnote{Notice that this momentum is superplanckian. This is not a problem, $p^{signal}_{V}$ is a coordinate dependent quantity and can be larger than the cutoff of our theory.} \cite{Maldacena:2017axo}
\begin{eqnarray}
\label{eq:uncertaintyPrinc}
p^{signal}_{V} \gtrsim \frac{1}{\Delta V} \approx \frac{\ell}{G_N h {\cal{A}}} \,, \label{uncertainty}
\end{eqnarray}
where $p^{signal}_{V}$ is the momentum of {\textit{one}} signal. Now imagine sending $N$ non-interacting signals. Then, $p_V = N p_V^{signal}$. Combining the uncertainty principle with the probe approximation condition \eqref{eq:probeAppShocks}, we find the following bound on the number of bits one can send through the wormhole,
\begin{eqnarray}
N \lesssim h {\cal{A}}\frac{r_{h}}{\ell} \,.
\end{eqnarray}
Note that we can send a large number of bits through the wormhole if we consider large black holes with $r_{h}\gg\ell$. However, this number is still much less than the theoretical maximum, which should scale with the entropy of the black hole,
\begin{equation}
	S_{BH}\approx \frac{r_{h}}{G_{N}}\gg \frac{r_{h}}{\ell}\,.
\end{equation}
From now on, we set $h={\cal{A}}=1$, so our results are correct up to order-one (small) numbers that depend on the details of the non-local interactions.
In \cite{Caceres:2018ehr}, it was shown that one way of increasing $N$ is to add rotation to the black hole. However, this is not enough to parametrically increase the amount of information transferred from order $r_h/\ell$ to the much larger $r_h/G_N$. 
An alternative way is to non-locally couple a large number of fields. This was done in the case of $AdS_2$ in \cite{Maldacena:2017axo}. Here we would like to analyze the consequences of this second approach in the case of $AdS_3$. Notice that if we interpret the non-local coupled fields as playing the role of the classical messages in the usual teleportation protocol, it is natural to consider many such fields to send more information. 

Following  \cite{Maldacena:2017axo}, we consider a deformation of the theory in which we couple $K$ fields
\begin{eqnarray}
\delta H = -\sum_{i=1}^{K} \ell \int d\phi \, h(t,\phi) \, {\cal{O}}^{i}_R (t,\phi) {\cal{O}}^{i}_L (-t,\phi) \,.
\end{eqnarray}
Assuming that the non-locally coupled fields are not interacting, the negative energy scales linearly with $K$. 

Large $K$ also allows us to enter the semiclassical regime. In order to couple the metric to the expectation value of the stress tensor, we would like the fluctuations in the stress tensor to be small compared to its mean. The fluctuations in the stress tensor depend on the scale, increasing at short distances. We would at least like the fluctuations to be small compared to the mean at scales of order the AdS radius. In the presence of $K$ light fields, the fluctuations in the stress tensor are of order
\be \label{eq:TUVVar}
\langle (\Delta T_U^{\ V})^2 \rangle \sim {K \over \ell^6} \,,
\ee
where we have focussed on the crucial component of the stress tensor for our analysis, $T_U^{\ V}$.
The mean value is\footnote{Notice the different scaling in $h$ between eq.\ \eqref{eq:TUVExpValue} and eq.\ \eqref{eq:TUVVar}. The expectation value of $T_{U}^{\ V}$ is zero in the absence of non-local coupling and hence it is proportional to $h$. The fluctuation of $T_{U}^{\ V}$ instead is non-zero also in the absence of the non-local coupling, due to the quantum fluctuation of the fields, and hence, to leading order, it is independent of $h$.}
\be \label{eq:TUVExpValue}
\langle T_U^{\ V} \rangle \sim {h K \over \ell^3} \,.
\ee
Imposing that the fluctuations are small compared to the mean gives
\be
h^2 K \gg 1~.
\ee
Since we require $h \ll 1$ in order to work to leading order in the source, we certainly need many fields, 
\be
K \gg {1 \over h^2} \gg 1 \,,
\ee
in order for the semiclassical description to be valid.

The opening of the wormhole is increased due to the increased negative energy,
\begin{equation}
	\Delta\!V \approx \frac{KG_{N}}{\ell}\,.
\end{equation}
As we have seen in section \ref{nonlocalint}, the non-local coupling is most effective at $t=0$ when the wormhole is shortest. For definiteness, we can then consider $K$ instantaneous non-local couplings all turned on at $t=0$. Turning on the coupling for a longer time would just modify the specific value of $\cal{A}$. The resulting picture is that of $K$ superimposed negative energy shocks. Notice that the probe approximation condition \eqref{eq:probeAppShocks} is not modified by the factor $K$. In the particle collision picture this means that we need to treat the collision between the $N$ signals and the $K$ shocks, in every thermal cell, as $K$ independent processes. However, the presence of many fields does influence the uncertainty principle condition because the opening of the wormhole increases with the available amount of negative energy
\begin{equation}
	p^{signal}_{V}\gtrsim\frac{1}{K}\frac{\ell}{G_{N}}\,.
\end{equation}
Combining this condition with \eqref{eq:probeAppShocks}, we again find a bound on the number of particles that can traverse the wormhole,
\begin{equation}
\label{no.ofpart}
	N\lesssim K\frac{r_{h}}{\ell}\,.
\end{equation}
This seem to suggest that we can send as much information as we want, if we allow $K$ to be large enough. However, the black hole has finite entropy and cannot be used to extract infinite amount of entanglement, so we expect a restriction for large enough values of $K$. 
For example, it is known that the presence of many species lowers the cutoff of the theory \cite{Dvali:2007wp,Kaloper:2015jcz}
\begin{equation}
	\ell_{UV}\gtrsim K G_{N}\,.
\end{equation}
The BTZ geometry cannot be treated as a semiclassical geometry when the UV cutoff becomes of the same order of the curvature scale, \textit{i.e.}\ we should have $\ell_{UV}\ll \ell$. This leads to an upper bound on $K$, 
\begin{equation}
	K\lesssim\frac{\ell}{G_{N}}\,. \label{Kbound}
\end{equation}
It turns out that this is already enough to make the bound on the number of particles consistent with the finiteness of the black hole entropy.
For the maximum value of $K$, we thus obtain
\begin{equation}
	N\lesssim\frac{r_{h}}{G_{N}}\approx S_{BH}\,. \label{eq_final_bound}
\end{equation}

Note that this requires the number of light operators in the CFT to be of order of the central charge $c$, and this is not the case in the usual known examples of $AdS_3/CFT_2$. Nevertheless, the GJW protocol seems to be robust enough that it will continue to make sense even in more exotic settings with a large number of light fields, where the UV cutoff of the field theory is not well-separated from the AdS scale, although the semiclassical bulk description will receive larger corrections.


\subsection{A multiple shocks bound}
\label{sec:multShockBound}
In this section, we will build the bulk geometry by gluing together black hole patches with different masses. This will be due to the effect of the non-local interaction and the message that will be modelled by shockwaves. By restricting the masses of the different patches to be positive, we will obtain constraints on the amount of energy that can be carried by the shockwaves. The main objective is to justify the species bound in equation (\ref{Kbound}) from a bulk perspective.

\begin{figure}
\centering
\begin{subfigure}[t]{.4\textwidth}
  \centering
  \includegraphics[height=6cm,keepaspectratio]{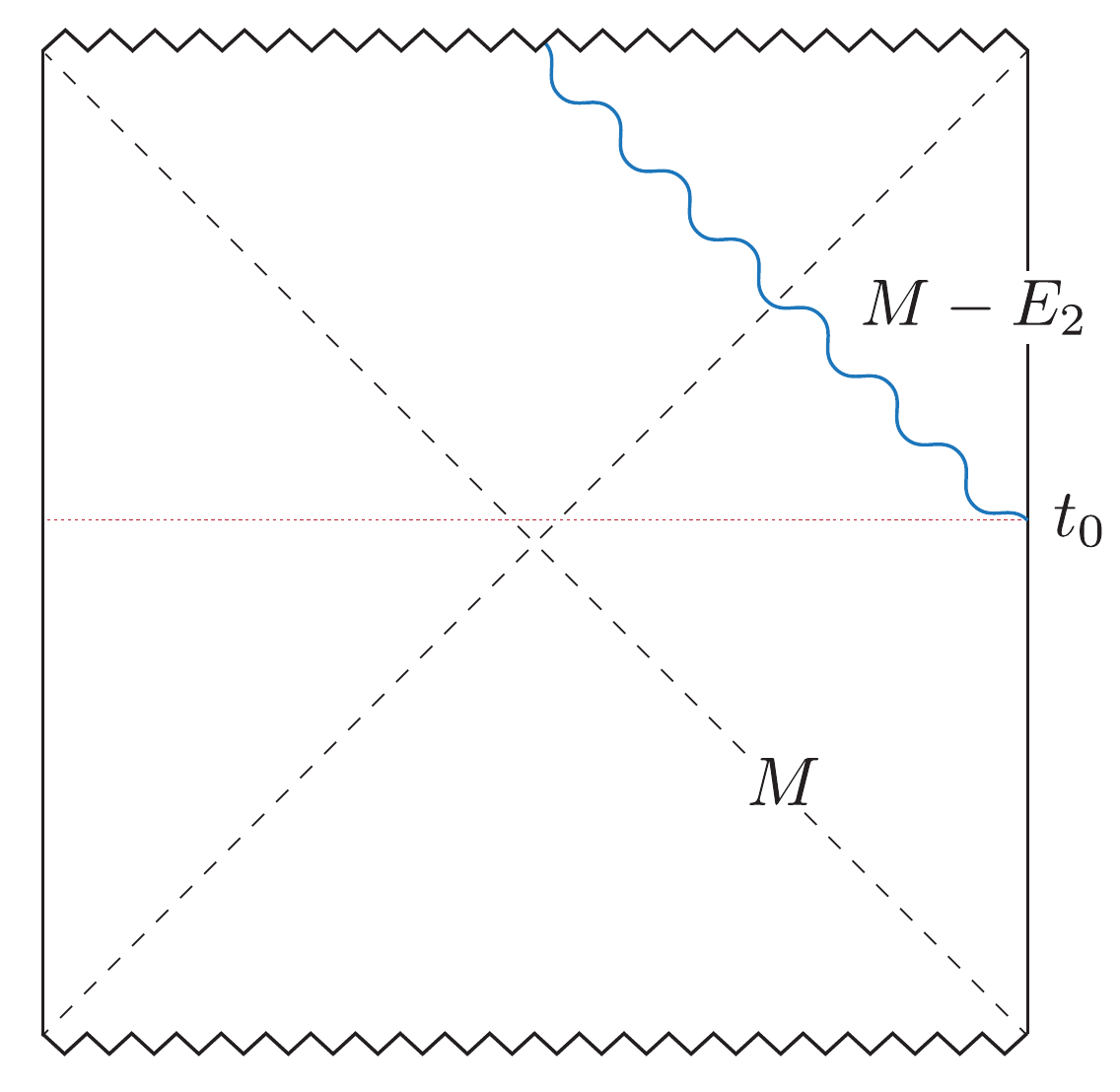}
  \caption{}
  \label{fig:sub11}
\end{subfigure}%
\begin{subfigure}[t]{.2\textwidth}
  \centering
  \includegraphics[height=6cm,keepaspectratio]{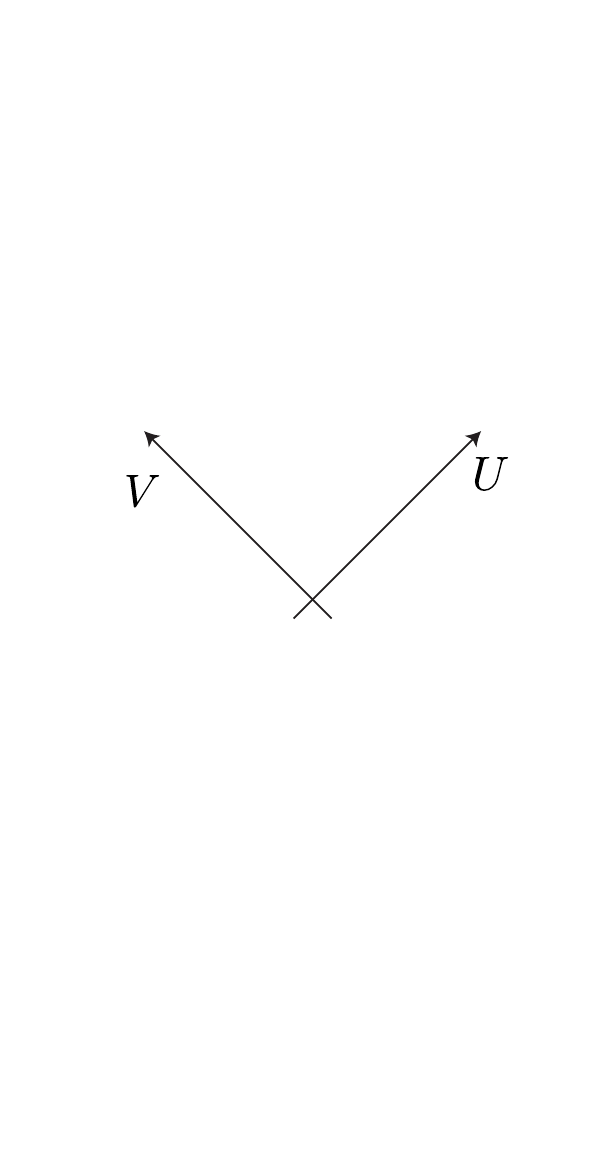}
  \caption*{}
  \label{fig:sub1222}
\end{subfigure}%
\begin{subfigure}[t]{.4\textwidth}
  \centering
  \includegraphics[height=6cm,keepaspectratio]{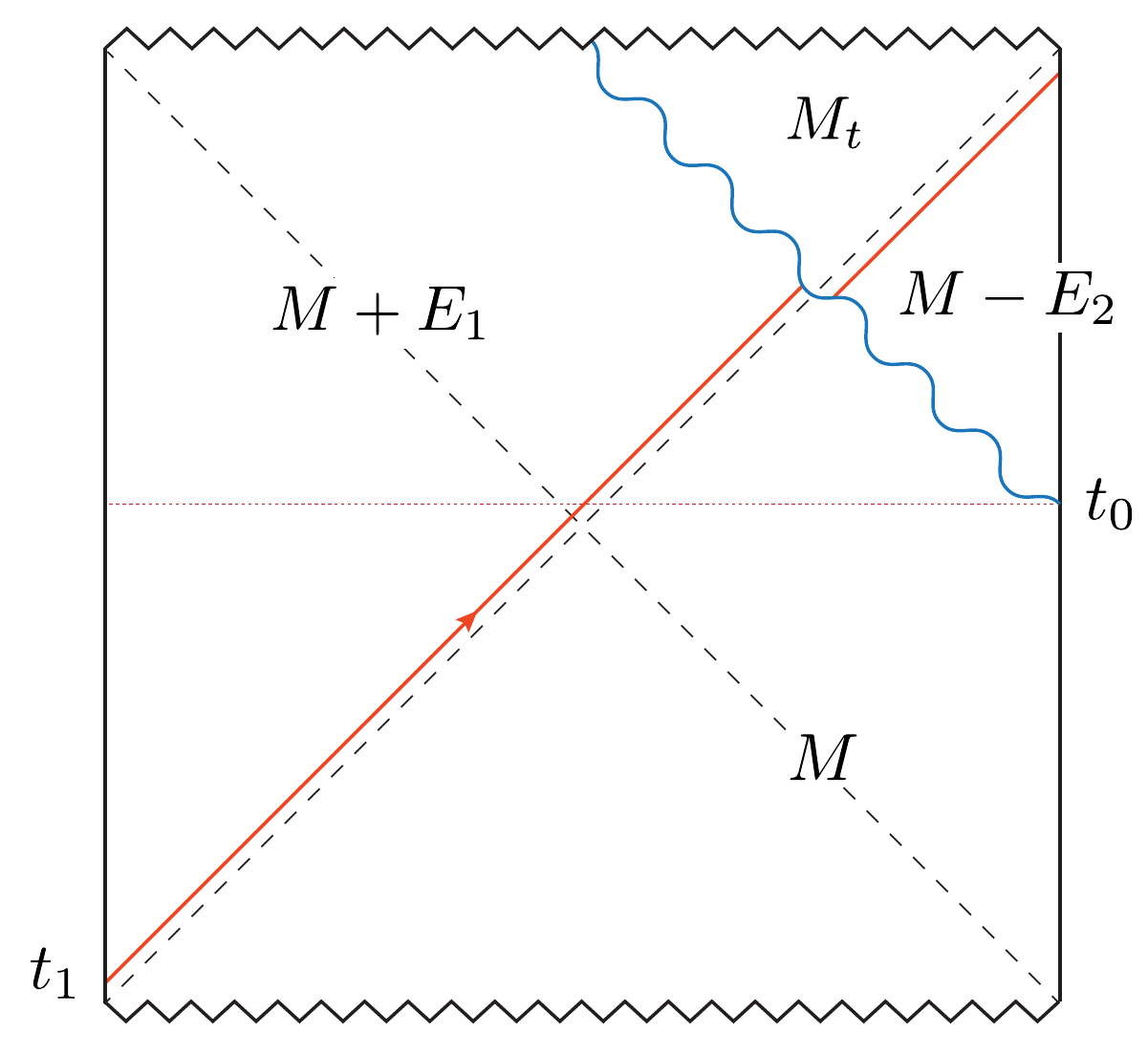}
  \caption{}
  \label{fig:sub21111}
\end{subfigure}
  \caption{Penrose diagrams of the different shockwave geometries. In (a), we only consider the effect of a negative energy shock of energy $-E_2$ (blue curvy line) sent at $t_0$. In (b), we add the effect of a second shock with positive energy $E_1$ sent at $t_1$ from the left boundary (solid red line). The resulting geometry is formed by gluing four AdS black hole patches with different masses.}
  \label{fig:Mald}
\end{figure}

Let's assume that the negative energy interaction between the two boundaries can be modeled in the bulk by the insertion of two negative-energy shockwaves at times $t_R=-t_L=t_0$. The message we would like to send through the wormhole can also be modeled as a shockwave, but now a positive-energy one. As shown in Fig.\ \ref{fig:Mald}, for the sake of analyzing whether the geometry becomes traversable, it is possible to neglect the effect of the left negative shockwave and consider the collision between two spherical shells, one with positive and the other with negative energy. The positive shell has energy $E_1$ and the negative one $-E_2$. As shown in \cite{Shenker:2013yza}, this gravitational problem involving the collision of two shocks can be solved by gluing different black-hole geometries together. Note that in this section it will be more convenient to use Schwarzschild coordinates and energies, and to only translate the final results into the null coordinates we used in previous sections.

First, let us consider the simpler example where there is only one negative shockwave -- see Fig.\ref{fig:sub11}. It is easy to see that if the mass of the original black hole is $M$ and the shock carries energy -$E_2$, then one should glue the geometry of a black hole with mass $M$ in the past of the shock with another one with mass $M-E_2$ at the future of the shock. 

The first bound comes from requiring that the mass of the second black holes geometry remains positive, \ie $M-E_2 >0$. The total energy $E_2$ of the negative shock is composed by the energy of the $K$ species. In each thermal cell, the energy should be of order of the local temperature and given that there are $r_h/\ell$ thermal cells, the total energy is given by $E_2 =\displaystyle K \beta^{-1} \frac{r_h}{\ell}=\displaystyle \frac{K \,r_h^2}{\ell^3}$. Using that the mass of the BTZ blackhole is $M \approx r_h^2/(G_N \ell^2)$, it is immediate to note that
\begin{equation}
K \lesssim \frac{\ell}{G_N} \,,
\end{equation}
that is exactly the species bound that appears in equation (\ref{Kbound}).

\

We can now add the positive energy shock, the message, and see if this setup provides additional constraints. In the case of two shocks colliding, the gluing gets more intricate as there are four different regions to consider. In \cite{Shenker:2013yza}, it was showed that it is enough to impose two gluing conditions in order to get a consistent answer: a continuity condition on the radius of the circle across the collision and a DTR regularity condition \cite{Dray:1985yt,redmount1985blue}. These two conditions allow us to find the mass of the black-hole in the post collision regime, $M_t$. See Fig. \ref{fig:sub21111}.

If we want to glue different metrics of the form $ds^2= -f_i(r) dt^2 + f_i(r)^{-1} dr^2 + r^2 d\phi^2$, from the DTR condition, we have that at the r-coordinate of collision $r_c$,
\begin{equation}
\label{DTR}
f_t(r_c)f_b(r_c)=f_l(r_c)f_r(r_c) \,,
\end{equation}
where $t,b,l,r$ stand for the top, bottom, left and right regions respectively. The difference with \cite{Shenker:2013yza} resides in that in our case one of the shocks carries negative energy (and is being sent at boundary time $t_0 \approx 0$). In this case, equation \eqref{DTR} becomes 
\begin{equation*}
\label{DTR2}
\left(r_c^2-8G_N M\ell^2\right)\left(r_c^2-8G_N M_t\ell^2\right)=\left(r_c^2-8G_N (M+E_1)\ell^2\right)\left(r_c^2-8G_N (M-E_2)\ell^2\right) \,.
\end{equation*}
This is sufficient to get $M_t$ as a function of the initial data. Moreover, if we want to write it as a function of the boundary times at which the shocks are emitted, we can translate $r_c$ in terms of the Kruskal coordinates. In the limit of small energies, $E_{1,2}/M\ll 1$, this can be done in any quadrant so for simplicity we consider the bottom one. In there, the horizon radius is the unperturbed one, given by $r_h^2=8G_N M \ell^2$. The $U$-coordinate of the negative shock wave is $U_-=e^{r_h t_0/\ell^2}$ and the $V$-coordinate of the positive shock wave is $V_+=e^{-r_h t_1/\ell^2}$. So, from equation (\ref{krus}), the collision radius becomes
\begin{equation}
r_c=r_h\dfrac{1-\exp\!\left(\dfrac{r_h}{\ell^2}(t_0-t_1)\right)}{1+\exp\!\left(\dfrac{r_h}{\ell^2}(t_0-t_1)\right)} \,.
\end{equation}
Plugging that back in equation \eqref{DTR2} is enough to find the mass of the black hole in the top region,
\begin{equation}
M_t=M+E_1-E_2-\frac{E_1E_2}{M}\cosh^2\left(\frac{r_h(t_0-t_1)}{2\ell^2}\right) \,.
\end{equation}

Note that for $(t_0-t_1)$ large enough, the last term grows exponentially leading to a negative mass in the upper region.\footnote{In fact, in three dimensions, this will happen even before the mass gets negative, as the BTZ black hole has a lower bound for its mass.} So, in the limit $E_{1,2}/M \ll 1$, imposing that $M_t$ should be positive results in 
\begin{equation}
M^2 \gtrsim E_1 E_2 \cosh^2\left(\frac{r_h(t_0-t_1)}{2\ell^2}\right) \,.
\end{equation}
We are interested in the case where $t_0 \approx 0$. Using that $e^{\frac{2r_h t_1}{\ell^{2}}}=-\displaystyle\frac{U_+}{V_+}$, that $U_+ V_+=-1$ on the boundary and that the shock with positive energy propagates close to the horizon $V_+=0$, we find that
\begin{equation}
M^2 \gtrsim \frac{E_1 E_2}{V_+} +\mathcal{O}(V_+) \,. \label{bound_shocks}
\end{equation}
For comparison, it is convenient to express this bound in terms of the center-of-mass energy of the collision,
\begin{equation}
	s = \frac{E_{1}E_{2}}{V_{+}}\frac{\ell^{2}}{r_{h}^{2}}\,.
\end{equation}
We obtain that 
\begin{equation}
	\label{eq:multipleBound}
	G_{N}\sqrt{s} \lesssim \frac{r_{h}}{\ell}\,
\end{equation}
where we used the definition of the black hole mass $M \approx r_{h}^{2}/G_{N}\ell^{2}$. Note that $s$ here corresponds to the collision between all the $N$ signals and the $K$ negative shocks. To compare with the previous bounds, we translate this expression into light-cone coordinates, where the stress-energy tensor for the message is already given in eq. (\ref{tvv}) and the one for the negative shocks has the generic form,
\begin{equation}		
T_{UU}=\frac{q_{U}}{r_h}\delta(U-U_0)\, .
\end{equation}
We have seen in the previous section that the magnitude of $T_{UU}$ scales as $\ell^{-1}$, see eq. \eqref{eq:negativeTUU}. So, given, that the negative shock is composed by $K$ signals, we expect $q_{U}=\displaystyle\frac{K r_h}{\ell}$. In light-cone coordinates, the center-of-mass energy squared is just
\begin{equation}
s=\frac{p_{V} \, q_{U} }{\ell^2},
\end{equation} 
and so, eq. (\ref{eq:multipleBound}) becomes a bound on $p_V$,
\begin{equation}
p_V \lesssim \frac{r_h \, \ell}{G_N^2 K} \,
\end{equation} 
that, combined with the uncertainty principle, gives yet another bound on the number of signals that can go through the wormhole,
\begin{equation}
N\lesssim \frac{r_h}{G_N}.
\end{equation}

Note that the final result is independent of $K$ so it would seem to imply that we can saturate the entropy bound without the need to couple many fields. However, note that while it is true that by solving the junction condition we have solved the full nonlinear Einstein equations, the same is not true for the field theory computation. As we explained before, the amount of negative energy generally decrease when we take into account the backreaction of the signal. Therefore, when we go beyond the probe approximation, we cannot treat the negative energy shock as a particle with a well defined momentum, $q_{U}$, which is independent of the signal momentum. In other words our previous computation implicitly assumed the validity of the probe approximation and the final result is only valid when the probe approximation is satisfied. 

\subsection{Beyond spherical symmetry}
So far we have bounded the amount on information we can transfer through the wormhole in the s-channel. We have seen that to send something through the wormhole we need $r_{h}\gg \ell$. However, it can be quite inconvenient to send signals spread over all the horizon. For example, a cat would have a hard time in such a delivery system. In this section we generalize our bound to signals that are localized to some region of size $b$ along the horizon. We begin by rederiving the probe approximation condition \eqref{eq:probeApp2} from a particle scattering perspective. 

As before, we approximate the interaction between the signal and the negative energy as a gravitational scattering between particles. Following \cite{Shenker:2014cwa} the condition for the validity of the probe approximation is given by 
\begin{equation}
	S_{cl}=\frac{1}{2}\int{d^{3}x\sqrt{-g}h_{UU}T^{UU}}\ll1\,,
\end{equation}
where $S_{cl}$ is the gravitational action evaluated on the shockwave geometry. Here $T^{UU}$ is the stress-energy tensor generated by the signal, and $h_{UU}$ is the gravitational field generated by the negative energy. We approximate the stress energy tensors of the signal and the negative energy respectively with \begin{equation}		
	T_{VV}=\frac{p_{V}}{r_{h}}\delta(V)t_{V}(\theta)\,;\quad T_{UU}=\frac{1}{\ell}\delta(U-U_{0})\,. \label{def stress}
\end{equation}
The first expression is the usual stress tensor generated by an energy shock with momentum $p_{V}$, where the transverse profile function $t_{V}(\theta)$ is a function with support on an interval of size $b/r_{h}$ and that integrates to 1. For our purpose, it will be enough to consider a step function. To define the second expression we have used that for the GJW construction the negative energy stress tensor scales like $\ell^{-1}$, as shown in \eqref{eq:negativeTUU}. The gravitational field obeys the following equation
\begin{equation}
	\left(-\partial_{\theta}^{2}+\frac{r_{h}^{2}}{\ell^{2}}\right)h_{UU}=G_{N}r_{h}^{2}T_{UU}\,,
\end{equation}
Since we are interested in the limit where $b\lesssim\ell\ll r_{h}$, we can approximate the horizon with an infinite line. This allows to avoid dealing with periodic boundary conditions and images. In other words let $x=\theta r_{h}/\ell$ we have 
\begin{equation}
	\left(-\partial_{x}^{2}+1\right)h_{UU}=G_{N}\ell^{2}T_{UU}\,,
\end{equation}
where $x$ takes values on the real line. The Green function for this equation is given by\footnote{The correct expression on the circle is given by $g(\theta-\theta')=\frac{\ell}{2r_{h}}\sum_{n\in \mathcal{Z}}\exp{\left(-\frac{r_{h}}{\ell}\left\vert \theta-\theta'+2\pi n\right\vert\right)}\,,$
we see that for $r_{h}\gg\ell$ we can neglect the images, \textit{i.e.}\ the terms with $n\neq0$. }
\begin{equation}
	\label{eq:greenFunction}
	g(x-x')= \frac{1}{2}e^{-\left\vert x-x'\right\vert}\,.
\end{equation}
Notice that this tells us that the gravitational interaction effectively shuts down when $\Delta\theta\approx \ell/r_{h}$, which is the angle corresponding to one thermal cell in the BTZ geometry. This means that, as we already pointed out earlier, the scattering between the messages and the negative energy shock naturally splits in $r_{h}/\ell$ independent shocks. In our case $T_{UU}$ does not depend on $\theta$ and so,
\begin{equation}
	h_{UU}(\theta)=G_{N}\ell^{2}\int{dx' g(x-x')T_{UU}}\approx G_{N}\ell\delta(U)\,.
\end{equation}
Finally, we find that the probe approximation is now given by 
\begin{equation}
	\label{eq:probeAppSpinning}
	S_{cl}\approx\frac{G_{N}p_{V}}{\ell}\ll1\,.
\end{equation}

Alternatively we can derive this expression by computing the time delay generated by the localized shock and impose that it is small. For the dependence on the transverse direction we take a simple step function
\begin{equation}
	t_{V}(\theta)=
	\begin{dcases}
		\frac{r_{h}}{b}\quad &0<\theta<\frac{b}{r_{h}}\,,\\
		0 &\text{otherwise}\,.
	\end{dcases}
\end{equation}
The gravitational field generated by this shock is given by 
\begin{equation}
	h_{VV}(x)=\frac{G_{N} \ell^{2} p_{V}}{b}\delta(V)\int_{0}^{b/\ell}{dx' e^{-\lvert x-x'\rvert}}\,.
\end{equation}
The integral above can be easily evaluated
\begin{equation}
	\int_{0}^{b/\ell}{dx' e^{-\lvert x-x'\rvert}}=
	\begin{dcases}
		e^{x}-e^{x-b/\ell}\quad &x<0\,,\\
		2-e^{-x}-e^{x-b/\ell}\quad &0\leq x\leq b/\ell\,,\\
		e^{b/\ell -x}-e^{-x}\quad &x>b/\ell\,.\\
	\end{dcases}
\end{equation}
Note that it is approximatively equal to $b/\ell$ in the interval $x\in[0,b/\ell]$, outside this interval is given approximatively by $b/\ell\,e^{-\vert x\rvert}$. We see that this quantity is  exponentially suppressed outside the thermal cell, \textit{i.e.}\ for $\vert x\rvert\gtrsim 1$. We conclude that 
\begin{equation}
	h_{VV}(\theta)\approx 
	\begin{cases}
		G_{N}\ell p_{V}\delta(V)\quad &\text{in the thermal cell}\,,\\
		0 &\text{otherwise}\,.
	\end{cases}
\end{equation}
From this we can find the time delay given by the positive shock on the negative shock
\begin{equation}
	\Delta U=\frac{1}{\ell^{2}}\int{dV h_{VV}}\approx 
	\begin{dcases}
	\frac{G_{N}p_{V}}{\ell}\quad &\text{in the thermal cell}\,,\\
		0 &\text{otherwise.}
	\end{dcases}
\end{equation}
If we require $\Delta U\ll1$ we recover \eqref{eq:probeAppSpinning}. 

Notice that, for a given value of momentum $p_{V}$, this is a more stringent requirement than the one we found for s-waves, \eqref{eq:probeApp}. Indeed, as we localize the message on shorter scales the energy density corresponding to a given value of $p_{V}$ increases, so it is natural that the probe approximation is harder to satisfy. However, the increased density ceases to play a role once we localize the signal on sub-AdS scales, \textit{i.e.}\ inside a thermal cell. This can be explained by looking at the Green function \eqref{eq:greenFunction}. This is free of divergences in the $x\rightarrow x'$ limit. In fact, it is approximately constant over the whole thermal cell. This means that when we localize the message on sub-AdS scale, the gravitational field generated is smeared over the whole thermal cell and it is always approximatively given by $G_{N}\ell p_{V}$. This is true independently of the value of $b$. Notice that this is special to three dimensions. Later, we will see that in higher dimensions the situation is not as simple. 

We would like now to proceed similarly to the previous subsection and bound $p_{V}$ from below. However, there is a complication. To localize the message along the horizon we need to excite higher angular momentum modes. Compared to s-waves, these modes have a harder time crossing the wormhole. Even after having emerged from the horizon thanks to the negative energy shock, they still need to overcome the potential barrier of the black hole. This provides an extra lower bound on the momentum needed by the signal, which for high enough angular momenta overcomes the one provided by the uncertainty principle. 

To find this new bound we consider the equation for spinning geodesics in the Schwarzschild metric. This can be obtained from the action
 \begin{equation}
	I=\frac{1}{2}\int{d\lambda\, g_{\mu\nu}\frac{dx^{\mu}}{d\lambda}\frac{dx^{\nu}}{d\lambda}}=\frac{1}{2}\int{d\lambda\left(-f(r)\dot{t}^{2}+\frac{\dot{r}^{2}}{f(r)}+r^{2}\dot{\phi}^{2}\right)}\,,
\end{equation}
where the dot represents derivatives with respect to $\lambda$.
The symmetries of the geometry ensure that along geodesics the energy, $p_{t}= -f\dot{t}\equiv-E$, and the angular momentum, $p_{\phi}=r^{2}\dot{\phi}\equiv L$, are conserved. We are interested in highly boosted particles, whose geodesics are approximately null. The equation can be obtained  by simply imposing $ds^{2}=0$, which is equivalent to the equation of motion of a particle in a one dimensional potential
\begin{equation}
	\dot{r}^{2}+V(r)=E^{2}\,;\quad V(r)\equiv\frac{L^{2}}{\ell^{2}}\left(1-\left(\frac{r_{h}}{r}\right)^{2}\right)\,.
\end{equation}
It is easy to see that the geodesic reaches the boundary only if 
\begin{equation}
\label{eq:EtoInf}
	E>\frac{L}{\ell}\,.
\end{equation}
To estimate the angular momentum needed localize the message to a region of size $b$ consider a Gaussian wave-packet
\begin{equation}
	f_{V}(\phi)\propto \exp\!\left(-\frac{(\phi-\phi_{i})^{2}}{b/r_{h}}\right)\,.
\end{equation}
Fourier transforming this expression it is easy to see that the needed angular momenta are those with $L\lesssim r_{h}/b$.

To compare this requirement with the uncertainty principle we first need to convert it to Kruskal coordinates. The momenta in Kruskal coordinates are given by 
\begin{equation}
\begin{split}
	p_{U}&=-\frac{E\ell^{2}}{2r_{h}U}+\frac{2Vr_{h}}{(1+UV)^{2}}\sqrt{\frac{E^{2}}{f^{2}}-\frac{L^{2}}{r^{2}f}}\,,\\
	p_{V}&=+\frac{E\ell^{2}}{2r_{h}V}+\frac{2Ur_{h}}{(1+UV)^{2}}\sqrt{\frac{E^{2}}{f^{2}}-\frac{L^{2}}{r^{2}f}}\,.
\end{split}
\end{equation}
To obtain this expression one first needs to find $p_{r}$ by imposing $p^{2}=0$. We are interested in the limit where $U\approx 1$ and $V\approx \Delta V\ll1$, for which we can approximate 
\begin{equation}
	f(U,V)=\frac{r_{h}^{2}}{\ell^{2}}\left(-\frac{4UV}{(1+UV)^{2}}\right)\approx -4\frac{r_{h}^{2}}{\ell^{2}}\Delta\!V\,.
\end{equation}
Using this we see that $p_{U}\approx 0$ while the $V$ component of the momentum is approximately given by
\begin{equation}
	 p_{V}\approx -\frac{E\ell^{2}}{r_{h}\Delta\!V}\,.
\end{equation} 
We conclude that the minimum required momentum needed to overcome the potential barrier is given by 
\begin{equation}
	\label{eq:spinMinBound}
	p_{V} \gtrsim \frac{\ell}{b}\frac{1}{\Delta V}\,.
\end{equation}
We see that for messages localized on scales smaller than a thermal cell this overcomes the uncertainty principle requirement \eqref{eq:uncertaintyPrinc}. Combining this with the probe approximation bound \eqref{eq:probeAppSpinning} we find
\begin{equation}
	N(b) \lesssim K\frac{b}{\ell}\,, \quad b \lesssim \ell\,.
\end{equation}
This in particular means that to send one single message localized to a region of size $b\ll \ell$ we need to couple $K\approx \ell/b$. We see that while it becomes increasingly difficult to send messages localized on sub-AdS scales, it is still possible if we are willing to couple a large number of fields. If we set $N=1$ in the above expression we can find the minimum allowed value of $b$ for a given $K$
\begin{equation}
	b(K) \gtrsim \frac{\ell}{K}\,.
\end{equation}
However, as we increase $K$ the cutoff of the theory gets lowered to $\ell_{UV}\approx KG_{N}$ and we should also impose that $b(K)>\ell_{UV}$. These two requirements coincide for $K_{min}\approx \sqrt{\ell/G_{N}}$, which leads to the following estimate for the minimum possible value of $b$
\begin{equation}
	b_{min}\approx \sqrt{\ell G_{N}}\,.
\end{equation}
Before concluding notice that in the opposite limit, $b\approx\ell$, the uncertainty principle and the potential barrier give the same bound and we find 
\begin{equation}
 N(b) \lesssim K\,, \quad b\gtrsim \ell\,.
\end{equation}
This means that instead of sending $K r_{h}/\ell$ s-waves we can also send messages localized in a thermal cell, sending $K$ such messages per thermal cell. 

\subsection{Comparison to Quantum Information Bounds}
In the previous sections, we have estimated the maximum information that can be sent through the wormhole with a bulk analysis. Here, we would like to briefly compare to a boundary analysis. A detailed boundary calculation is difficult because the theory is strongly coupled, but we can still place bounds on the amount of information transferred. 
As explained in  \cite{Maldacena:2017axo}, we can think of the procedure as quantum teleportation.\footnote{We refer to quantum teleportation in a broad sense, without specializing to a particular quantum communication protocol. For a more detailed description of specific quantum communication protocols that might be dual to the traversable wormhole see \cite{Bao:2018msr}.} In standard quantum teleportation, if Alice and Bob share and EPR pair, they can use it as a resource to transfer a qubit. One qubit can be transferred at the cost of using up one entangled EPR pair and sending two bits of classical information. Here we think of the left-right coupling as playing the role of the classical communication, as explained in more detail in \cite{Maldacena:2017axo}.
 
With these identifications, the amount of information sent is bounded by the decrease in entanglement entropy between the two CFT's,\footnote{We thank David Berenstein for discussions on this point.}
\begin{equation}
	N \lesssim -\Delta S_{EE} \,. \label{prev_equation}
\end{equation}
We can compute the change in entropy from the change in energy induced by the non-local coupling
\begin{equation}
	\Delta S_{EE} = \beta \Delta E\,.
\end{equation}
This equation is valid because in the thermofield double the entanglement entropy between the two sides is equal to the thermal entropy. This statement is clear before acting with the coupling. After acting with the coupling, a bulk calculation tells us that the entanglement entropy is still equal to the thermal entropy, because the bulk geometry is still an eternal AdS-Schwarzschild black hole. We do not have a direct CFT argument for this equality.

As mentioned above the boundary theory is strongly coupled, so computing $\Delta E$ directly on the boundary would be hard. The best we can do is to assume that the result of such a computation would match with the one we obtained with the bulk analysis. In other words we assume that the change in energy is given by one thermal quantum per coupling and per thermal cell,
\begin{equation}
	\Delta E \approx-\frac{K\ell}{\beta^{2}}\,,
\end{equation}
where the number of thermal cells is given by $\ell/\beta$. This is the only input we need from the bulk computation in this section. It would be interesting to check this, at least for some examples of weakly coupled boundary theories, see for example \cite{Berenstein:2019yfv}. 

Combining the above equations we find 
\begin{equation}
	N \lesssim \frac{K\ell}{\beta}\,,
\end{equation}
which is $K$ bits per thermal cell. This agrees with our bulk estimate found in equation \eqref{no.ofpart} by requiring the bulk geometry to remain in the probe approximation. Clearly the entanglement entropy cannot decrease below zero, so an absolute bound is
 \begin{equation}
N \lesssim S_{EE}\,.
 \end{equation}
This absolute bound is saturated (up to order one prefactors) when we maximize the number of species providing the negative energy, as given in equation \eqref{eq_final_bound}.

Notice that since we interpret the non-local coupled fields as playing the role of the classical messages in the usual teleportation protocol, it is natural to consider many such fields to send more information. However, from the CFT point of view it does not seem necessary that these are couplings between \textit{different} fields, it might be possible to couple the same field but at different times. This seems to suggest that also in the bulk, if we were able to go beyond the probe approximation, we might be able to send order $S_{BH}$ bits without coupling a parametrically large number of fields. 

\subsection{Generalization to $d+1$ dimensions} \label{sec:higherdim}
The picture we uncovered in the previous section is rather simple. Signals need to be highly boosted to fall through the wormhole, \eqref{eq:uncertaintyPrinc}. Their backreaction on the geometry modifies the non-local coupling configuration, generally inducing a reduction of the negative energy. In other words, they close the wormhole. We showed this at linear level, but the analysis of \cite{Caceres:2018ehr,Maldacena:2017axo} suggests that this is true also at nonlinear level. Certainly the negative energy is preserved if we can neglect the backreaction of the signal altogether, \textit{i.e.}\ the probe approximation is valid, \eqref{eq:probeApp}. The combination of this bound with the requirement that every signal is boosted enough constraints the amount of information that can be sent through the wormhole, see \eqref{no.ofpart}. 

In this section we would like to understand how the bound on information transfer is modified in $d+1$ dimensions. We expect the above picture to still be valid. Namely, the amount of information that can be transferred is bounded by a combination of the probe approximation and the uncertainty principle. Unfortunately, it is hard to carry out explicitly the calculation of GJW in higher dimensions, in particular we cannot find an expression for the negative energy. We will assume that the stress-energy tensor generated by the non-local coupling still scales with the AdS radius,
\begin{equation}
\label{eq:higherTUU}
	T_{UU}\propto \frac{K}{\ell^{d-1}}\,,
\end{equation}
where we have already included $K$ species.

To find the equivalent of the probe approximation bound \eqref{eq:probeApp} in $d+1$ dimensions, we again impose that the time delay generated by the positive energy shock is small, $\Delta U\ll1$. The relation between the time delay and the stress energy tensor of the shock in general dimensions is still given by \eqref{eq:timeDelayT}, where the Newton constant is now related to the Planck length by $8\pi G_{N}=\ell_{P}^{d-1}$. In terms of the total momentum carried by the signal the condition \eqref{eq:probeApp} becomes 
\begin{equation}
\label{eq:higherProbe1}
	\Delta U = \frac{G_{N}p_{V}}{r_{h}^{d-1}}\ll1\,.
\end{equation}
We can again rewrite this condition in terms of coordinate independent quantities if we model the interaction between the signal and the negative energy as gravitational scattering. We showed that the gravitational interaction, close to the horizon of the BTZ black hole, is localized to a thermal cell of size $\ell$ and therefore, the collision could be split in independent events, $K$ for each thermal cell. We demanded that each of these collision events was well described in the probe approximation, \eqref{probewiths}.
In higher dimensions it is still true that the gravitational scattering is localized to a thermal cell. The limit on the validity of the probe approximation for gravitational scattering in $d+1$ dimensions, see for example  \cite{Giddings:2007bw}, is given by 
\begin{equation}
	\label{eq:higherProbe2}
	G_{N}s_{cell}\frac{1}{\ell^{d-3}}\ll1\,,
\end{equation}
Here we have set the impact parameter to be of order $\ell$. To find $s_{cell}$, first notice that, as can be seen from \eqref{eq:higherTUU}, the negative energy particles still carry one unit of momentum per thermal cell. To find the momentum carried by the positive energy shock per thermal cell, we simply divide the total momentum $p_{V}$ by the number of thermal cells, $r_{h}^{d-1}/\ell^{d-1}$. Under this identifications it easy to see that \eqref{eq:higherProbe2} agrees with \eqref{eq:higherProbe1}. 

The uncertainty principle condition carries on to higher dimension without modifications. However, the size of the wormhole opening is now given by 
\begin{equation}
	\Delta V=G_{N}\int{T_{UU}}\approx K\frac{G_{N}}{\ell^{d-1}}\,.
\end{equation}
We can combine the uncertainty principle with the probe approximation requirement to bound the amount of information we can send through the wormhole. The computation is identical to the one above, the final result is 
\begin{equation}
	N \lesssim K\left(\frac{r_{h}}{\ell}\right)^{d-1}\,.
\end{equation}
Similarly to the lower dimensional case the amount of information we can transfer scales with $K$ and the number of thermal cells. 
We can find a bound on $K$ recalling that in higher dimension the UV cutoff is renormalized as follows \cite{Dvali:2007wp,Kaloper:2015jcz}
\begin{equation}
	\ell_{UV} \gtrsim \left(KG_{N}\right)^{\frac{1}{d-1}}\,.
\end{equation}
Taking this into account we see that for the maximal value allowed, $K\approx \ell^{d-1}/G_{N}$, we find 
\begin{equation}
	N_{max}\approx \frac{r_{h}^{d-1}}{G_{N}}\approx S_{BH}\,.
\end{equation}
This is the generalization to higher dimensions of our bound for information transfer in the s-wave channel. We see that in any dimensions, for the maximum value of $K$ allowed by the species bound, we saturate the black-hole entropy. 

We now turn to the case of localized messages presented above. For messages localized in regions larger than the AdS scale, everything works the same as in the three dimensional case. Instead of sending s-waves, we can send signals localized to thermal cells, $K$ such signals per thermal cell. This is because, as pointed out above, the gravitational propagator in higher dimensions also decays exponentially outside the thermal cell. The behaviour at shorter distances, instead, is qualitatively different in higher dimensions. The propagator is not constant inside the thermal cell but acquires a singularity in the $x'\rightarrow x$ limit
\begin{equation}
	g(x-x')\propto \frac{1}{\lvert x-x'\rvert^{d-3}}\,.
\end{equation}
Here $x_{i}$ are dimensionless coordinates defined similarly to \eqref{eq:greenFunction}. As a consequence the gravitational field generated by signals localized on sub-AdS scales is not constant anymore inside the thermal cell and the analysis is not as simple. In particular, it might be possible to try and send many localized messages per thermal cell. We will not make this computation, but we simply imagine sending many localized messages superimposed at the center of the thermal cell. The time delay $\Delta U$ has now a non-trivial profile inside the thermal cell. For simplicity we impose that the maximum value of this delay inside the thermal cell is small. In principle we would need to solve 
\begin{equation}
	\left(-\partial_{\Omega}^{2}+\frac{r_{h}^{2}}{\ell^{2}}\right)h_{VV}=G_{N}r_{h}^{2}T_{VV}\,,
\end{equation}
where $\partial_{\Omega}^{2}$ is  the Laplacian on the $(d-1)$-dimensional transverse sphere. Inside the thermal cell we can neglect the second term in the parenthesis and approximate the sphere with a plane. The equation reduces to the Poisson equation for a Newtonian potential in $d-1$ dimensions. For radii larger than the $b/\ell$, the solution is simply given by\footnote{Notice that this equation is only valid for $d>2$. Moreover, in the case $d=3$ the polynomial reduces to a logarithm, $\lvert x\rvert^{3-d}\rightarrow \log{\lvert x\rvert}$.} 
\begin{equation}
	h_{VV}\approx \frac{p_{V}\ell^{d-3}}{\lvert x\rvert^{d-3}}\delta(V)\,.
\end{equation}
The maximum is given by $\lvert x\rvert=b/\ell$, which leads to 
\begin{equation}
	\Delta U  \lesssim \frac{G_{N}}{\ell^{2}b^{d-3}}p_{V}\ll1\,.
\end{equation}
Similarly to the three dimensional case the signal needs to be boosted enough to overcome the angular momentum potential barrier, see \eqref{eq:spinMinBound}.
Combining these requirements we can find that the information transfer bound is given by\footnote{The correct result for $d=3$ is $N(b) \lesssim K\frac{b}{\ell}\frac{1}{\log{\ell /b}} $}
\begin{equation}
	N(b) \lesssim K\left(\frac{b}{\ell}\right)^{d-2}\,.
\end{equation}
As compared to the three dimensional case it is indeed harder to send localized signals, for a given value of $K$. We can find the minimum value of $b$ for a given $K$ by setting $N=1$ in the above equation 
\begin{equation}
	b(K) \gtrsim K^{\frac{-1}{d-2}}\ell\,.
\end{equation}
As we increase the number of coupling we renormalize the UV cutoff of the theory, so we need also to check that $b$ is larger than $\ell_{UV}$. We have 
\begin{equation}
	b(K) \gtrsim K^{\frac{1}{d-1}}\ell_{P}\,.
\end{equation}
Combining these two bounds we find that the minimum possible value of $b$ is given by\footnote{In the $d=3$ this formula is correct up to logarithmic corrections.}
\begin{equation}
	b_{min}=\left(\frac{\ell}{\ell_{P}}\right)^{\frac{1}{d-2}}\ell_{P}\,.
\end{equation}
We conclude that in high enough dimensions we can localize messages on scales smaller than what is possible in the three dimensional case. The reason is that even though it is harder to localize messages for a fixed value of $K$, in higher dimensions we can couple more fields before the UV cutoff reaches the AdS scale. 

\section{Discussion and future directions} \label{sec_conc}

In this work, we computed bounds on the amount of information that can be transferred in the traversable wormhole construction by Gao, Jafferis and Wall (and slight generalizations of it). This computation was motivated by some seemingly problematic features of the GJW wormhole. Namely, the perturbative nature of the non-local coupling opened the wormhole in the bulk only at a sub-Planckian scale, making it dubious whether large amounts of information could be actually transferred before closing the wormhole again. On the other hand, the wormhole is built perturbatively around the eternal blackhole geometry. This means that the entanglement entropy between the two boundaries is large and given by the black hole entropy. From the boundary perspective, and assuming this protocol is somehow dual to teleportation, this generates, in principle, a large amount of entropy available to teleport information from one side to the other. The question then becomes clear: is there a way we can use all that amount of entanglement entropy to maximize the information transfer through the wormhole?

In section \ref{secGJW}, we studied in detail the construction of GJW, allowing for different types of non-local sources and computing the amount of negative energy generated by each of them. In particular, we found a simple analytic formula for the case where the sources are instantaneously turned on, avoiding much of the numerical computation that are usually done in the literature. 

In section \ref{secBounds}, we found that the amount of information transferred in the standard GJW wormhole is of order ${\cal{O}} (h \, r_h/\ell)$. Note that this, in general, is much smaller than the entanglement entropy,  $r_h/G_N$. 

Nevertheless, we found that large black holes can allow for more than one bit of information transfer. This contrast with previous results in the literature, in particular with \cite{Hirano:2019ugo}, where it is claimed that the maximum amount of information that can be transferred is ${\cal{O}}(1)$. The difference lies in that their construction only couples the s-wave between two boundaries, while the coupling we consider is local in space and therefore couples many angular modes. Coupling only the s-wave  leads to a smaller amount of negative energy, allowing for at most one bit of information to be transmitted, as the authors find.  Moreover the particular infinite boost limit that is taken in \cite{Hirano:2019ugo} does not seem physically well motivated: in the infinite boost limit, the collision between the negative energy and the signal will have an arbitrarily high center of mass energy and not be well-described in the semiclassical regime.

We also showed that it is possible to increase the amount of bits that can go from one boundary to the other by introducing a large number $K$ of light fields coupled between the two boundaries. Using a combination of bounds coming from the uncertainty principle, the probe approximation and the existence of many species, we found that in principle it would be possible to send $N \approx S_{BH}$ bits of information. This is interesting, since it maximizes the amount of information that can be sent, at least from the boundary teleportation perspective.

The results on this paper rely on several assumptions and approximations of the traversable wormhole geometry. It will be interesting to further relax these assumptions and see whether it is possible to improve on our results. In the following, we comment on interesting possible future directions.

\

\noindent {\textit{Beyond the probe approximation.}} Most of the results presented rely in the so-called probe approximation, assuming that the scattering processes between the shocks are small. This seems to be a strong restriction because it is only possible to send the maximum possible amount of information by allowing an extraordinarily large number $K$ of light bulk fields, $K \approx \ell^{d-1}/G_N$. This large number of light fields lowers the UV cutoff of the bulk theory. Also, many holographic theories do not have a large number of light fields.

It would interesting to see if it is possible to saturate the amount of information transferred without the need of so many fields by going beyond the probe approximation. The calculation we presented in the multiple shocks section \ref{sec:multShockBound} is in this spirit, showing that independently of $K$, we get a bound on $N$ coming from the gluing of the multiple shocks geometries. 

One issue in going beyond the probe approximation is that the backreaction of the signal on the geometry means that we would have to re-compute the stress tensor coming from the coupled quantum fields, because we can no longer use the propagator in the BTZ background in calculating the stress tensor.  This was explored in \cite{Caceres:2018ehr}, where it was claimed that going beyond the probe approximation just reduces the amount of negative energy generated, and therefore does not allow for more information transfer.

We are not fully convinced by these results for the following reason. The effect of the backreaction is to create a time delay in the propagation across the signal. The new bulk-to-boundary propagator can be computed in the presence of the signal. Effectively, the signal induces a relative shift between the left and right boundary times. In the absence of the signal, the most effective boundary-boundary coupling occurs when the left and right boundary points are both halfway up the Penrose diagram, at $t_L = t_R =0$, or at points related to this by symmetry. Upon introducing the signal, the shift means that the most effective coupling occurs when both points are in the lower half of the Penrose diagram. However, \cite{Caceres:2018ehr} does not consider allowing the coupled boundary points to be in this region. It would be very interesting (and probably not difficult for the authors of \cite{Caceres:2018ehr}) to extend their analysis into this regime.

If it is in fact possible, this would be rather surprising, given that from the teleportation picture it would seem that we would either need to couple a large number $K$ of different fields, or we would need to keep the coupling turned on for a long time. 

To summarize: in this paper we have calculated how much information can be sent while remaining in the probe regime, where the signal does not disturb the leading order calculation of the negative energy due to coupled quantum fields. These probe regime calculations and arguments are reliable. However, we do not have a persuasive bulk argument explaining why the information that can be sent is bounded by these probe regime calculations. It seems feasible to carry the analysis beyond the probe regime in the future.

\

\noindent {\textit{Quantum metric fluctuations.}} Since the wormhole is open for such a short time, shorter than the Planck time, one might worry that quantum metric fluctuations will have a large correction on the transmission of a semiclassical message. We postpone a more complete discussion of these quantum fluctuations to future work. However, at least in 2+1 dimensions, we can argue that the quantum fluctuations will have a small effect. Quantum fluctuations include two effects: the thermofield double state includes a superposition of different black hole masses, and the black holes can be decorated by boundary gravitons. 

Due the special properties of 2+1 dimensions, all of these metrics can be thought of as BTZ black holes, deformed arbitrarily close to the boundary by gravitons.\footnote{We thank Jan de Boer for discussion on this point.} In analyzing the signal, these effects can all by combined into an uncertainty in the dimensionless time, $t/\beta$, that the signal is emitted from the left boundary or received by the right boundary. The effects of these perturbations are suppressed by powers of the gravitational coupling; we believe that the quantum uncertainty is given by
\be
\Delta (t  / \beta ) \approx \sqrt{G_N \over \ell} \,.
\ee
Since we are interested in sending signals whose time duration is just a bit less than the thermal scale, these quantum corrections to the width of the signal  are neglibible.

\

\noindent {\textit{Beyond perturbative calculations.}} Many of the confusions that arise in the context of GJW are due to the perturbative nature of the non-local interaction. It would be interesting to find solutions at finite coupling and/or construct eternal wormholes in this context. In two bulk dimensions, it is possible to create an eternal wormhole \cite{Maldacena:2018lmt}, but the generalization to higher dimensions is not as straightforward --see, for instance \cite{Freivogel:2019lej}.

\

\noindent {\textit{Beyond three spacetime dimensions.}}
The GJW construction relies heavily on the simplicity of the BTZ correlators. In section \ref{sec:higherdim}, we provide plausible generalizations to general dimensions of the bounds on information found on this work. It would be desirable to find a framework in higher dimensions where these claims could be checked by explicit calculations.

\

\noindent {\textit{Beyond black hole horizons.}} A natural framework to study traversable wormholes are horizons in de Sitter spacetimes. Due to the nature of the cosmological horizon, the insertion of shockwaves naturally provides a mechanism for traversable wormholes. In the context of two dimensional gravity, it is possible to glue cosmological horizons in the IR, with an AdS boundary in the UV, and construct such shockwave solutions \cite{Anninos:2018svg}. The nice feature about those solutions is that they do not need the insertion of non-local, negative energy couplings. It would be interesting to see whether they can be generalized to higher dimensions and compare the maximum bounds on information transferred in each case.

We hope to come back to some of these ideas in a future communication.

\section*{Acknowledgements}

We gratefully acknowledge discussions with David Berenstein, Elena Caceres, Alejandra Castro, Shira Chapman, Jan de Boer, Daniel Jafferis, Anderson Misobuchi, and Juan Pedraza. BF, DN, and AR are supported by the ERC Consolidator Grant QUANTIVIOL. This work is part of the $\Delta$ ITP consortium, a program of the NWO that is funded by the Dutch Ministry of Education, Culture and Science (OCW).

\bibliographystyle{JHEP}
\bibliography{bibliography}

\end{document}